\documentclass[aps,superscriptaddress,showpacs,showkeys,nofootinbib,twocolumn,floatfix]{revtex4}
\usepackage{graphicx,epsfig,wrapfig,color}
\usepackage{color}

\definecolor{darkgreen}{rgb}{0,0.65,0}

\newcommand{\be}{\begin{equation}}
\newcommand{\ee}{\end{equation}}
\newcommand{\ba}{\begin{eqnarray}}
\newcommand{\ea}{\end{eqnarray}}
\newcommand{\di}{\!{\rm d}}
\newcommand{\la}{\langle}
\newcommand{\ra}{\rangle}
\begin{document}
\newcommand*{\Yale}{Department of Physics Yale University, 
   New Haven, CT 06511-8499, U.S.A.}\affiliation{\Yale}
\newcommand*{\UConn}{
   Department of Physics, University of Connecticut,
   Storrs, CT 06269-3046, U.S.A.}\affiliation{\UConn}
\title{\boldmath
	Energy momentum tensor, stability, and the $D$-term of $Q$-balls}
\author{Manuel Mai}\affiliation{\Yale}
\author{Peter Schweitzer}\affiliation{\UConn}
\date{May 2012}
\begin{abstract}
We study the energy-momentum tensor of stable, meta-stable and unstable 
$Q$-balls in scalar field theories with U(1) symmetry. 
We calculate properties such as charge, mass, mean square radii and the 
constant $d_1$ (``$D$-term'') as functions of the phase space angular 
velocity $\omega$. We discuss the limits when $\omega$ approaches 
the boundaries of the region in which solutions exist, and derive
analytical results for the quantities in these limits.
The central result of this work is the rigorous proof that $d_1$ is 
strictly negative for all finite energy solutions in the $Q$-ball system.
We also show that for $Q$-balls stability is a sufficient, but not 
necessary, condition for $d_1$ to be negative.
\end{abstract}
%
%
%
\maketitle

\section{Introduction}
\label{Sec-1:introduction}

The matrix elements of energy momentum tensor $T_{\mu\nu}$ (EMT) 
\cite{Pagels} contain basic information, such as mass 
\cite{Ji:1994av} and spin of the particle \cite{Ji:1996ek}, 
and the constant $d_1$. This constant denotes the value of the 
corresponding form factor at zero-momentum transfer, and is the 
only experimentally unknown ``charge'' associated with the EMT. 
Its physical meaning was uncovered in \cite{Polyakov:2002yz} where it 
was shown that $d_1$ is related to the spatial distribution of internal 
forces. This is analog to the interpretation of the electric form factor 
as the Fourier transform of the electric charge distribution \cite{Sachs}
(and subject to the same type of limitations \cite{Sachs,Ji:1991ff} 
due to relativistic corrections).

The EMT form factors can in principle be studied through generalized 
parton distribution functions \cite{Muller:1998fv,Ji:1998pc} accessible 
in hard exclusive reactions such as deeply virtual Compton scattering
\cite{Adloff:2001cn,Airapetian:2001yk,Stepanyan:2001sm,Munoz-Camacho:2006hx}.
In this work we shall loosely refer to $d_1$ as the $D$-term, although
they coincide strictly speaking only for asymptotically large 
renormalization scales \cite{Polyakov:1999gs}. 
Theoretical studies of EMT form factors were presented in chiral 
perturbation theory, lattice QCD, effective chiral field theories 
or models 
\cite{Polyakov:1999gs,Mathur:1999uf,Petrov:1998kf,Goeke:2007fp,Goeke:2007fq,
Cebulla:2007ei,Kim:new,Donoghue:1991qv,Megias:2004uj,Liuti:2005gi,Guzey:2005ba}.

It is a striking observation that in all theoretical studies $d_1$ 
was found negative  --- for pions, nucleons, nuclei. Results from 
chiral soliton models \cite{Goeke:2007fp,Goeke:2007fq,Cebulla:2007ei,Kim:new} 
suggest a connection of the sign of $d_1$ and stability of the particle.
The naturally emerging questions are: could the negative sign of $d_1$ 
be a model-independent feature, a theorem? And, what is the precise 
relation of $d_1$ and stability?

This work is devoted to the study of the EMT of $Q$-balls 
\cite{Friedberg:1976me,Coleman:1985ki} with the aim to shed 
further light on the sign of $d_1$ and its connection to stability.
$Q$-balls are non-topological solitons in theories with global Abelian 
\cite{Coleman:1985ki} or non-Abelian \cite{Safian:1987pr} symmetries,
and have been discussed in a variety of approaches with a wide range
of applications in particle physics, cosmology, and astrophysics
\cite{Cohen:1986ct,Alford:1987vs,Lee:1991ax,Kusenko:1997ad,Multamaki:1999an,Kasuya:1999wu,Volkov:2002aj,Clougherty:2005qg,Clark:2005zc,Gleiser:2005iq,Schmid:2007dm,Gani:2007bx,Verbin:2007fa,Sakai:2007ft,Bowcock:2008dn,Tsumagari:2008bv,Arodz:2008jk,Gabadadze:2008sq,Campanelli:2009su}.

The $Q$-ball equations of motions admit stable, meta-stable and 
unstable solutions. 
This makes them an ideal ground for our purposes to explore the
connection of the stability of an object and its $D$-term.
In this work we will be interested in the ground state 
$Q$-ball solutions  \cite{Coleman:1985ki}. 
The equations of motions admit also radial excitations 
\cite{Volkov:2002aj} which will be subject to a separate work 
\cite{work-in-progress}.

The stability of $Q$-ball systems was studied in many works
\cite{Volkov:2002aj,Gleiser:2005iq,Gani:2007bx,Sakai:2007ft,Tsumagari:2008bv}.
But this is, to the best of our knowledge, the first time this issue is 
addressed from the point of view of the EMT, and the $D$-term.
In particular, we present the first rigorous proof in a dynamical
system that $d_1$ must be negative, and clarify the connection
of $d_1$ and stability. The outline of this work is as follows.

In Sec.~\ref{Sec-2:EMT-and-Qballs}, after a brief review of $Q$-balls, 
we derive the expressions for the energy 
density $T_{00}(r)$, pressure and shear forces, $p(r)$ and $s(r)$,
related to the stress tensor $T_{ik}(r)$, and prove analytically that 
exact solutions of the $Q$-ball equations of motions satisfy the stability 
condition which is a consequence of the conservation of the EMT.
We show that the stability condition is equivalent to the 
virial theorem, as it is the case also in soliton models of the
nucleon \cite{Goeke:2007fp,Goeke:2007fq,Cebulla:2007ei}.

In Sec.~\ref{Sec-3:ground-states} we study $Q$-ball
properties such as charge, mass, mean square radii, and the 
$D$-term in a chosen potential as functions 
of the angular velocity $\omega$ in the U(1)-space in the
region $\omega_{\rm min} < \omega < \omega_{\rm max}$ in which
the equations of motion admit finite energy solutions.
An interesting observation is that among the quantities we study
$d_1$ varies most strongly with $\omega$.

In Sec.~\ref{Sec-4:limit-omega-to-min} and \ref{Sec-5:limit-omega-to-max}
we then focus on the behavior of the $Q$-ball properties as 
$\omega$ approaches the boundaries of the region in which finite
energy solutions exist, i.e.\ the limits $\omega\to\omega_{\rm min}$ 
and $\omega\to\omega_{\rm max}$ respectively. We derive in both cases 
analytical results which describe the behavior the different 
quantities in these limits and which are fully supported by the numerical
results.

In Sec.~\ref{Sec-6:the-proof} we formulate two independent
proofs that $d_1$ of $Q$-balls is strictly negative, and clarify the 
relation to stability. We show that stability is a sufficient but not 
necessary condition for $d_1$ to be negative.

The Sec.~\ref{Sec-7:conclusions} contains a summary of our findings 
and the conclusions. Some technical details and supplementary results
are discussed in the Appendix.

\newpage
\section{\boldmath  $Q$-balls }
\label{Sec-2:EMT-and-Qballs}

We consider the relativistic field theory \cite{Coleman:1985ki}
of a complex scalar field $\Phi(x)$ defined by the Lagrangian
\be\label{Eq:Lagrangian}
   {\cal L} = \frac12\,(\partial_\mu\Phi^\ast)(\partial^\mu\Phi) 
            - V\,.
\ee
The potential $V\ge 0$ is defined in terms of the positive 
constants $A$, $B$, $C$ (with $4AC>B^2$ to guarantee $V>0$
for $\phi\neq 0$) as follows 
\be\label{Eq:potential}
   V = A\,(\Phi^\ast\Phi)-B\,(\Phi^\ast\Phi)^2+C\,(\Phi^\ast\Phi)^3\:.
\ee
The Lagrangian is invariant under global 
U(1) symmetry transformations $\Phi\to\Phi\,e^{i\alpha}$, 
$\Phi^\ast\to\Phi^\ast e^{-i\alpha}$ with $\alpha$ real.
This system admits non-topological soliton solutions 
\cite{Coleman:1985ki} which in the soliton rest frame are given by
\be\label{Eq:ansatz}
   \Phi(t,\vec{x}) = \exp(i\omega t)\,\phi(r)\;,\;\;\;r=|\vec{x}\,|\:.
\ee
The Euler-Lagrange equations of the theory in (\ref{Eq:Lagrangian}) imply
for the radial field $\phi(r)$ the following differential equation
(here and in the following primes denote differentiation
with respect to the argument)
\be\label{Eq:eom}
   \phi^{\prime\prime}(r)+\frac2r\;\phi^\prime(r)+\omega^2\phi
   - V^\prime(\phi) = 0\:,
\ee
which is subject to the boundary conditions
\ba
&&  \phi(0) \equiv \phi_0 = {\rm const},   \;\;
    \phi^\prime(0) = 0,   \;\nonumber\\
&&   \phi(r)\to 0\;\;\mbox{for}\;\;r\to\infty\;.
\label{Eq:boundary-conditions}\ea

The Noether theorem applied to the global U(1) symmetry
implies the conserved charge 
\be\label{Eq:charge}
   Q = \int\di^3x\;\rho_{\rm ch}(r)\;,\;\;\;
   \rho_{\rm ch}(r) = \omega\;\phi(r)^2\,.
\ee
The sign of $\omega$ determines the sign of the charge $Q$.
In the following we assume $\omega>0$ without loss of generality.
The presence of a continuous global symmetry is essential for 
the existence of the soliton.
More precisely, finite energy solutions exist for $\omega$ in 
the range
\be\label{Eq:condition-for-existence}
      \omega_{\rm min}^2 < \omega^2 < \omega_{\rm max}^2 \;,\;\;\;\;\; \ee
with 
\ba
&&\;\;  \omega_{\rm min}^2 = \min\limits_\phi \biggl[\frac{2\,V(\phi)}{\phi^2}
        \biggr] = 2A\biggl(1-\frac{B^2}{4AC}\biggr)>0\;, \nonumber\\
&&\;\;  \omega_{\rm max}^2 = V^{\prime\prime}(\phi)\biggl|_{\phi=0}=2A\;.
        \label{Eq:condition-for-existence-2}\ea

From (\ref{Eq:eom},~\ref{Eq:boundary-conditions}) we obtain 
for $\phi(r)$ the following small- and large-$r$ behavior 
(the dots indicate subleading terms)
\ba
\label{Eq:asymp-small}
    \phi(r) &=& 
    \phi_0 + \biggl(V^\prime(\phi_0)-\omega^2\phi_0\biggr)\,\frac{r^2}{6} 
    + \dots  \;\;\mbox{small $r$},\;\;\;\;\\
    \phi(r)&=&\!\frac{c_\infty}{r}\;
    \exp\biggl(-r\sqrt{\omega_{\rm max}^2-\omega^2}\biggr)\, 
    + \dots  \;\mbox{large $r$}.
\label{Eq:asymp-large}
\ea
The constants $\phi_0$ and $c_\infty$ are known, of course, only after 
solving the boundary value problem (\ref{Eq:eom},~\ref{Eq:boundary-conditions}).

\subsection{Stability criteria}
\label{Sec-2a:stability-criteria}

Solutions for $\omega$ satisfying the existence condition  
(\ref{Eq:condition-for-existence}) can be classified as 
(a) stable, (b) meta-stable, and (c) unstable $Q$-balls, 
see e.g.\ \cite{Tsumagari:2008bv} for an overview.

(a) If $M$ denotes the mass of the soliton, and $m$ the mass of the
field $\Phi$, which is $m=\sqrt{2A}=\omega_{\rm max}$, then the
absolute stability condition can be expressed as \cite{Lee:1991ax}
\be\label{Eq:absolute-stability}
       M < m\,Q\,,\;\;\; m\equiv \omega_{\rm max}\;.
\ee

(b) Meta-stable solutions do not satisfy (\ref{Eq:absolute-stability})
but are stable with respect to small 
fluctuations, and satisfy a weaker ``classical stability condition'' 
\cite{Friedberg:1976me,Lee:1991ax} which can be formulated 
in the equivalent ways
\be\label{Eq:classical-stability}
        \frac{\di}{\di\omega}\left(\frac{M}{Q}\right) \ge 0
         \;\;\; \Leftrightarrow \;\;\;\frac{\di Q}{\di\omega} \le 0
         \;\;\; \Leftrightarrow \;\;\;\frac{\di^2 M}{\di Q^2} \le 0\;,
\ee
i.e.\ a critical $\omega_c$ (extreme charge $Q_c$)
exists at which the slope (curvature) of the quantities in 
(\ref{Eq:classical-stability}) changes.

(c) Solutions satisfying neither the stronger condition
(\ref{Eq:absolute-stability}) nor the weaker condition 
(\ref{Eq:classical-stability}) are unstable.

\subsection{\boldmath The EMT of $Q$-balls}
\label{Subsec-3a:EMT-and-Qballs}

For the theory defined by the Lagrangian (\ref{Eq:Lagrangian}) 
the canonical energy momentum tensor 
\be\label{Eq:EMT-general}
   T_{\mu\nu} 
  =\frac{\partial{\cal L}}{\partial(\partial^{\mu}\Phi     )}\,\partial_\nu\Phi
  +\frac{\partial{\cal L}}{\partial(\partial^{\mu}\Phi^\ast)}\,\partial_\nu\Phi^\ast
   - g_{\mu\nu} \,{\cal L}
\ee
is symmetric and static. The energy density, which defines the mass 
$M=\int\di^3x\,T_{00}$, is given by 
\be\label{Eq:T00}
   T_{00}(r) = \frac12\,\omega^2\phi(r)^2+\frac12\,\phi^\prime(r)^2+V(\phi)\;.
\ee
The $T_{0k}$ components vanish, i.e.\ the $Q$-ball has spin zero.
(Of course,  $Q$-ball solutions can be assigned a non-zero spin
by means of appropriate projection techniques \cite{Rajamaran}.)
Finally, the $T_{ij}$ components describe the stress tensor 
\be\label{Eq:Tik}
   T_{ij} = \biggl(\frac{x_ix_j}{r^2}-\frac13\,\delta_{ij}\biggr) s(r) 
   + \delta_{ij}\,p(r)
\ee
with the distribution of the shear forces, $s(r)$, and pressure, $p(r)$,
given by
\ba
    s(r) &=& \phi^\prime(r)^2 \label{Eq:shear}\\
    p(r) &=& \frac12\,\omega^2\phi(r)^2-\,\frac16\,\phi^\prime(r)^2 -V(\phi)
    \label{Eq:pressure}\;.
\ea
The dimensionless constant $d_1$ is defined through the stress
tensor $T_{ij}$ \cite{Polyakov:2002yz}, and can be expressed 
in terms of $s(r)$ and $p(r)$ 
(cf.\ Sec.~\ref{Subsec-3b:conservation-of-EMT-and-Qballs})
as follows
\ba\label{Eq:def-d1}
    d_1 &=& -\,\frac{4\pi}{3}\,M\int_0^\infty\di r\;r^4s(r)
    \label{Eq:def-d1-shear}\;,\\
        &=& \phantom{-}\,5\pi\,M\int_0^\infty\di r\;r^4p(r)
    \label{Eq:def-d1-pressure}\;.
\ea
The large-$r$ asymptotics (\ref{Eq:asymp-large})
ensures that all integrals (which define
$M$, $Q$, $d_1$, mean square radii, etc) exist.

\subsection{Consequences from conservation of EMT}
\label{Subsec-3b:conservation-of-EMT-and-Qballs}

For a static EMT $\partial^\mu T_{\mu\nu}=0$ is equivalent
to $\nabla^i T_{ij}=0$ which, using the decomposition (\ref{Eq:Tik}),
implies 
\cite{Polyakov:2002yz}
\be\label{Eq:diff-eq-s-p}
      \frac{2}{r}\,s(r) + \frac{2}{3}\,s^\prime(r) + p^\prime(r) = 0\;.
\ee
In order to prove that (\ref{Eq:diff-eq-s-p}) holds for $Q$-balls, we 
insert $s(r)$ and $p(r)$ from Eqs.~(\ref{Eq:shear},~\ref{Eq:pressure})
into (\ref{Eq:diff-eq-s-p}), which yields
\ba\label{Eq:diff-eq-s-p-proof}
\hspace{-5mm} && \frac{2}{r}\,s(r)+\frac{2}{3}\,s^\prime(r)+p^\prime(r)\\
\hspace{-5mm} && \;\;\; = \phi^\prime(r)
      \,\biggl(\phi^{\prime\prime}(r)+\frac2r\;\phi^\prime(r)
      +\omega^2\phi(r) - V^\prime(\phi)\biggr) = 0\nonumber
\ea
due to the equations of motion in  Eq.~(\ref{Eq:eom}).

From (\ref{Eq:diff-eq-s-p}) one can derive the equivalent
representations (\ref{Eq:def-d1-shear},~\ref{Eq:def-d1-pressure})
for $d_1$ in terms of $s(r)$ and $p(r)$, and other general relations 
\cite{Goeke:2007fp}. For instance, multiplying 
(\ref{Eq:diff-eq-s-p}) by $r^3$ and integrating (by parts) 
over $r$ from zero to infinity yields the ``stability condition''
(also referred to as ``von Laue--condition,'' see 
\cite{von-Laue} and references there in) 
\be\label{Eq:stability-condition}
      \int\limits_0^\infty \di r\;r^2p(r)=0\;.
\ee

In order to prove (\ref{Eq:stability-condition}) for $Q$-balls 
we integrate by parts
(primes denote derivatives with respect to the arguments, 
the finite upper integration limit $R$ is for later purposes) 
\be\label{Eq:proving-stability-1.1}
     \int\limits_0^R\di r\;r^2\,p(r) = 
     \Biggl[\,\frac{r^3}{3}\;p(r)\Biggr]_0^R
     -\int\limits_0^R\di r\;\frac{r^3}{3}\;p^\prime(r)
\ee
Next we notice that 
\ba\label{Eq:proving-stability-1.2}
     p^\prime(r) 
     &=&  \biggl(-\frac13\,\phi^{\prime\prime}(r)
         +\omega^2\phi(r)-V^\prime(\phi)\biggr)\,\phi^\prime(r)\nonumber\\
     &=& -\,\frac43\,\phi^\prime(r)\phi^{\prime\prime}(r)
         -\frac2r\;\phi^\prime(r)^2
     \nonumber\\
     &=& -\,\frac2{3r^3}\biggl[r^3\phi^\prime(r)^2\biggr]^{\!\prime}    
\ea
where we used 
$\omega^2\phi-V^\prime(\phi)=-\phi^{\prime\prime}(r)-\frac2r\;\phi^\prime(r)$
in the first step which holds due to the equations of motion (\ref{Eq:eom}).
Hence, using $s(r)=\phi^\prime(r)^2$, we obtain
\be\label{Eq:proving-stability-1.3}
     \int\limits_0^R\di r\;r^2\,p(r) = 
     \Biggl[\frac{\displaystyle r^3}{3}\;
     \biggl(p(r)+\frac23\;s(r)\biggr)\Biggr]_0^R \;.
\ee
The small- and large-$r$ behavior of the solutions in (\ref{Eq:asymp-small})
guarantees that the lower and (after taking $R\to\infty$) upper
integration limits in (\ref{Eq:proving-stability-1.3})
vanish which proves (\ref{Eq:stability-condition}).

It is instructive to prove (\ref{Eq:stability-condition}) independently as 
follows. Let $\phi(r)$ be a $Q$-ball solution with charge $Q$ and mass
$M$, which we rewrite by means of (\ref{Eq:charge},~\ref{Eq:T00}) in terms
of the ``charge,'' ``surface,'' and ``potential energies'' 
(where we leave the number of dimensions $D=3$ general)
\be\label{Eq:virial-theorem-I}
      M=\frac12\,E_{\rm ch}
       +\frac12\,E_{\rm surf}
       + E_{\rm pot}
\ee
\ba\label{Eq:virial-theorem-Ib}
     E_{\rm surf}=\int\di^Dx\,\phi^\prime(r)^2\;,&&
     E_{\rm pot }=\int\di^Dx\,V(\phi)\;,\nonumber\\
                I=\int\di^D x\,\phi(r)^2\;,&&
     E_{\rm ch}=\frac{Q^2}{I} \;. \label{Eq:def-Esurf-Ech-Epot}
\ea
We consider dilatational variations $\phi(r)\to\phi(\lambda r)$ 
of the solutions with a positive parameter $\lambda$.
Substituting in the integrals in (\ref{Eq:virial-theorem-Ib}) 
$\vec{x}\to\lambda \vec{x}$ yields
\be\label{Eq:virial-theorem-II}
      M(\lambda)=\frac12\,E_{\rm ch}\;\lambda^D
      +\frac12\,E_{\rm surf} \;\lambda^{2-D}
      +E_{\rm pot} \;\lambda^{-D} \;.
\ee
$M(\lambda)$ has a minimum at $\lambda=1$,
because $\phi(\lambda r)$ for $\lambda=1$ is a $Q$-ball 
solution which minimizes the energy functional.
We obtain, using again (\ref{Eq:charge}) and the
definitions in (\ref{Eq:virial-theorem-Ib}),
\ba\label{Eq:virial-theorem-III}
     0 &\stackrel{!}{=}& \frac1D\;
     \frac{\partial M(\lambda)}{\partial\lambda}\biggl|_{\lambda=1} 
     =  
       \frac{1}{2}\,E_{\rm ch}
      +\frac{2-D}{2D}\,E_{\rm surf}
      -E_{\rm pot} \nonumber\\
     &=& \!\!
     \int\di^Dx\,\biggl\{
       \frac{1}{2}\,\omega^2\phi(r)^2
      -\frac{D-2}{2\,D}\,\phi^\prime(r)^2
      - V(\phi)\biggr\} .\;\;\;\;\;\;\;
\ea
For $D=3$ we identify the expression (\ref{Eq:pressure})
for $p(r)$ in the curly brackets of (\ref{Eq:virial-theorem-III})
which completes our alternative proof of (\ref{Eq:stability-condition}). 
Eq.~(\ref{Eq:virial-theorem-III}) is known as virial theorem 
\cite{Kusenko:1997ad}.
Notice that (\ref{Eq:virial-theorem-III}) can be used to eliminate,
for instance, the potential energy term from (\ref{Eq:virial-theorem-I}),
leading to
\be\label{Eq:M-omegaQ-Esurf}
      M=\omega \,Q+\frac{1}{D}\,E_{\rm surf}\;.
\ee

As a last application of (\ref{Eq:diff-eq-s-p}) we integrate this
equation over $r$ from zero to infinity. This yields the relation
\cite{Goeke:2007fp}
\be\label{Eq:p(0)}
     p(0) = 2\int\limits_0^\infty\di r\;\frac{s(r)}{r}\,,
\ee
which provides a helpful cross check for numerical calculations, and 
implies the following interesting relation: inserting in (\ref{Eq:p(0)})
the expressions (\ref{Eq:shear},~\ref{Eq:pressure}) yields
\be\label{Eq:interesting-relation}
     \frac12\,\omega^2\phi_0^2 - V(\phi_0) = 
     2\int\limits_0^\infty\di r\;\frac{\phi^\prime(r)^2}{r}\,.
\ee
This is interesting, because the left-hand-side depends on $\phi_0$ 
only while the right-hand-side is a functional of $\phi^\prime(r)$ 
where $\phi_0$ drops out. Below in Sec.~\ref{Sec-3a:Ueff} we will discuss
the physical interpretation of (\ref{Eq:interesting-relation}).

\subsection{\boldmath
Relations among $Q$-ball properties}

Further interesting relations among different $Q$-ball properties
follow from combining 
(\ref{Eq:charge},~\ref{Eq:T00},~\ref{Eq:shear},~\ref{Eq:pressure}) as
\be\label{Eq:relation-T-p-rho-s}
     T_{00}(r)+p(r) = \omega\,\rho_{\rm ch}(r)+\frac13\,s(r)\;.
\ee
Integrating (\ref{Eq:relation-T-p-rho-s}) over  $\;\di^3x$ we recover 
(\ref{Eq:M-omegaQ-Esurf}) for $D=3$ (the derivations are equivalent,
but (\ref{Eq:M-omegaQ-Esurf}) elucidates the relation of the factor 
$\frac13$ to the dimensionality of the space).
Next, we define the ``surface tension'' $\gamma$ and the mean square
radius $\la r^2_s\ra$ of the shear forces $s(r)$ as follows
\ba\label{Eq:def-gamma-rs}
     \gamma = \int\limits_0^\infty\,\di r\;s(r) \; , \;\;\;
     \la r^2_s\ra=\frac{\int\limits_0^\infty\,\di r\;r^2 s(r)} 
                       {\int\limits_0^\infty\,\di r\;s(r)} \,.
\ea
Thus, the surface energy $E_{\rm surf} = \int\di^3x\,s(r)$, 
Eq.~(\ref{Eq:def-Esurf-Ech-Epot}), can be written as
$E_{\rm surf} = 4\pi\,\la r_s^2\ra\,\gamma$ which is what one
expects for a spherical object with a well-defined surface
and radius $\la r_s^2\ra^{1/2}$. 
In Sec.~\ref{Sec-4:limit-omega-to-min} we will see that these 
notions make sense for $Q$-balls in a certain limit.

Finally, we weight (\ref{Eq:relation-T-p-rho-s}) with $r^2$, and 
integrate over $\;\di^3x$. This allows to express $d_1$ in terms
of other properties as
\be\label{Eq:d1-from-properties}
    d_1 = \frac59\,\biggl(
    \omega\,Q\,M\,\la r^2_Q\ra - M^2 \, \la r^2_E\ra  \biggr)\,,
\ee
with the mean square radii of energy and charge densities
defined as
\ba\label{Eq:def-mean-square-radii-E-M}
     \la r^2_E\ra=\frac{\int\di^3x\;r^2\,T_{00}(r)}
                       {\int\di^3x\,     T_{00}(r)} \,,\;\;
     \la r^2_Q\ra=\frac{\int\di^3x\;r^2\,\rho_{\rm ch}(r)}
                       {\int\di^3x\,     \rho_{\rm ch}(r)}\,.\;\;
\ea

\subsection{Parameters and numerics}
\label{Sub-3e:numerics}

In our numerical study we fix the parameters as
\be\label{Eq:parameters}
         A=1.1\,,\;\;\;\;
         B=2.0\,,\;\;\;\;
         C=1.0\,.
\ee
(for which in \cite{Volkov:2002aj} radial $Q$-ball excitations were 
found; the latter originally motivated our study, but will be discussed 
in a separate work \cite{work-in-progress}).  
This yields the following range of allowed $\omega$-values
\be\label{Eq:omega-range}
         0.2 < \omega^2 < 2.2\,.
\ee
The parameter set $(A,\,B,\,C)$ could be assigned 
physical units, say (GeV$^2$, GeV$^0$, GeV$^{-2}$).
Then  $\omega$, $M$ would be given in GeV, mean 
square radii in GeV$^{-2}$, etc. But
for simplicity we will work with dimensionless quantities.

The numerical method is as follows.
For a given $\omega$ the differential equation 
(\ref{Eq:eom}) is solved
with slightly shifted initial conditions 
$\phi(\varepsilon)\equiv\phi_\varepsilon$ and
$\phi^\prime(\varepsilon)=\frac13(V^\prime(\phi_\varepsilon)-\omega^2
\phi_\varepsilon)\varepsilon$ with numerical parameters 
$\varepsilon$ in the range $10^{-10}$ to $10^{-4}$.
We checked that the results do not depend on $\varepsilon$.
Finite energy solution are found using the shooting method by varying 
the initial value $\phi_\varepsilon$ until $\phi(r)\to 0$ at large $r$.

The quality of the numerics is monitored by testing that 
{\sl (i)} the differential equation (\ref{Eq:diff-eq-s-p}) holds, 
{\sl (ii)} the stability condition (\ref{Eq:stability-condition}) is valid,
{\sl (iii)}  different expressions for $d_1$ in
(\ref{Eq:def-d1-shear},~\ref{Eq:def-d1-pressure},
\ref{Eq:d1-from-properties}) yield the same result, 
{\sl (iv)} the same value for $p(0)$ follows from 
(\ref{Eq:pressure},~\ref{Eq:p(0)}).
We find a relative numerical accuracy of 
${\cal O}(10^{-6})$ or better.

\section{Ground state $Q$-balls}
\label{Sec-3:ground-states}

In this Section we discuss the ground state properties of $Q$-balls
in our potential (\ref{Eq:potential}) for different values of $\omega$. 

\subsection{\boldmath 
Effective potential $U_{\rm eff}$ and $\phi_0$}
\label{Sec-3a:Ueff}

\begin{figure}[b!]
\centering
\includegraphics[width=7cm]{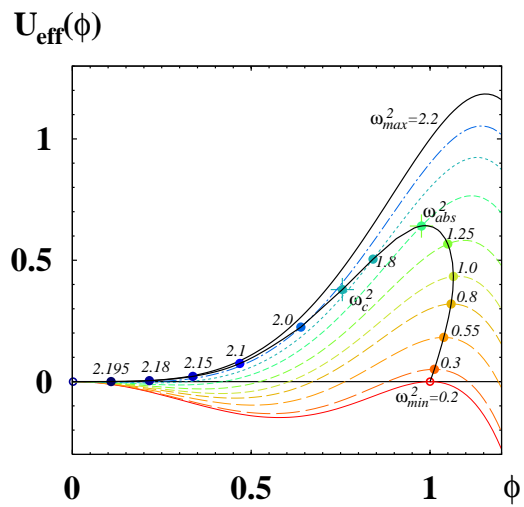}
\caption{\label{Fig-01:eff-pot}
The effective potentials $U_{\rm eff}(\phi)=\frac12\,\omega^2\phi^2-V(\phi)$
as functions of $\phi$ for selected values of $\omega^2$ in the range
(\ref{Eq:omega-range}). The circles show the initial values $\phi_0$
for each $\omega^2$, which lie on a curve starting and ending at 
the limiting values $\omega_{\rm min}^2=0.2$ and $\omega_{\rm max}^2=2.2$
(marked by open circles). The special values 
$\omega^2_{\rm c}\approx 1.9$ and $\omega_{\rm abs}^2\approx 1.55$
are discussed in text.}
\end{figure}

Identifying $r\to t$ and $\phi(r)\to x(t)$, the equation of motion 
(\ref{Eq:eom}) can be read \cite{Coleman:1985ki} 
as the Newtonian equation 
\ba
\label{Eq:Newtonian-eom}
&& \mbox{\it\"x}(t)=F_{\rm fric}-\nabla U_{\rm eff}(x) \\
&& F_{\rm fric}=-\frac2t\,\mbox{\it\. x}(t) \,,\;\;\; 
U_{\rm eff}=\frac12\omega^2\,x^2-V(x) \,.\nonumber
\ea
describing the motion of a particle of unit mass under the influence
of the time- and velocity-dependent friction $F_{\rm fric}$ in the 
effective potential $U_{\rm eff}$ shown in Fig.~\ref{Fig-01:eff-pot}.
The initial and boundary values (\ref{Eq:boundary-conditions}) 
mean that at $t=0$ the particle starts from the position $x_0$
with zero velocity, and comes to rest in the origin $x=0$ 
after infinite time. Thus $x(t)>0$ and the 
particle never stops at finite $t$. 
This implies decreasing monotony of the ground state fields, 
$\phi(r)>0$ and $\phi^\prime(r)<0$ for $0 < r < \infty$.

The pressure (\ref{Eq:pressure}) is given at 
$r=0$ by $p(0) = U_{\rm eff}(\phi_0)$ and the condition 
(\ref{Eq:condition-for-existence}) guarantees the existence 
of a region of $\phi$ with $U_{\rm eff}(\phi)>0$ \cite{Coleman:1985ki}.
This proves that
\be\label{Eq:pressure-positive-at-small-r} 
         p(0)>0 \;.
\ee

Now also the physical interpretation of (\ref{Eq:interesting-relation})
is evident. The left hand side of (\ref{Eq:interesting-relation}) is the 
initial potential energy. 
The right hand side of (\ref{Eq:interesting-relation}) is the work 
$W=\int F_{\rm fric}\,\di x$ the particle does to overcome 
the friction before coming to rest
at $x=0$ with zero effective potential energy. 

\newpage

Solutions exist for all $\omega$ in the range 
$\omega_{\rm min}<\omega<\omega_{\rm max}$.
With our numerical method, we were able to find solutions 
in the subinterval $0.216\le\omega^2\le 2.195$.

The effective potentials 
$U_{\rm eff}(\phi)=\frac12\,\omega^2\phi^2-V(\phi)$ 
are shown in Fig.~\ref{Fig-01:eff-pot}
for selected  $\omega$ including the limiting cases
$\omega_{\rm min}$ and $\omega_{\rm max}$. 
On each of the $U_{\rm eff}(\phi)$-curves
in Fig.~\ref{Fig-01:eff-pot} we marked the initial conditions 
$\phi_0$ which solve the boundary value problem 
(\ref{Eq:eom},~\ref{Eq:boundary-conditions}).
The $U_{\rm eff}(\phi_0)$ for different $\omega$ lie on a curve which 
exhibits a global maximum close to $\omega_{\rm abs}^2\approx 1.55$, and 
changes the curvature around $\omega_{\rm c}^2\approx 1.9$ 
(these frequencies will be discussed in detail in 
Sec.~\ref{Sec-3c:special-omega}).
This curve starts and ends at the limiting points
\be\label{Eq:phi-in-limiting-cases}
   \lim\limits_{\omega\to\omega_{\rm max}} \phi_0 = 0 \, , \;\;\;\; 
   \lim\limits_{\omega\to\omega_{\rm min}} \phi_0 =  
   \sqrt{\frac{B}{2C}}=1 \;,\ee
with the potential $U_{\rm eff}(\phi_0)\to0$ in both cases, 
see App.~\ref{App-A:trivial-solutions}.
$U_{\rm eff}(\phi_0)$ as function of $\phi_0$ is not unique
for $\phi_0 \ge 1$.

\begin{figure}[hb!]

\vspace{-4mm}

\includegraphics[width=8.25cm]{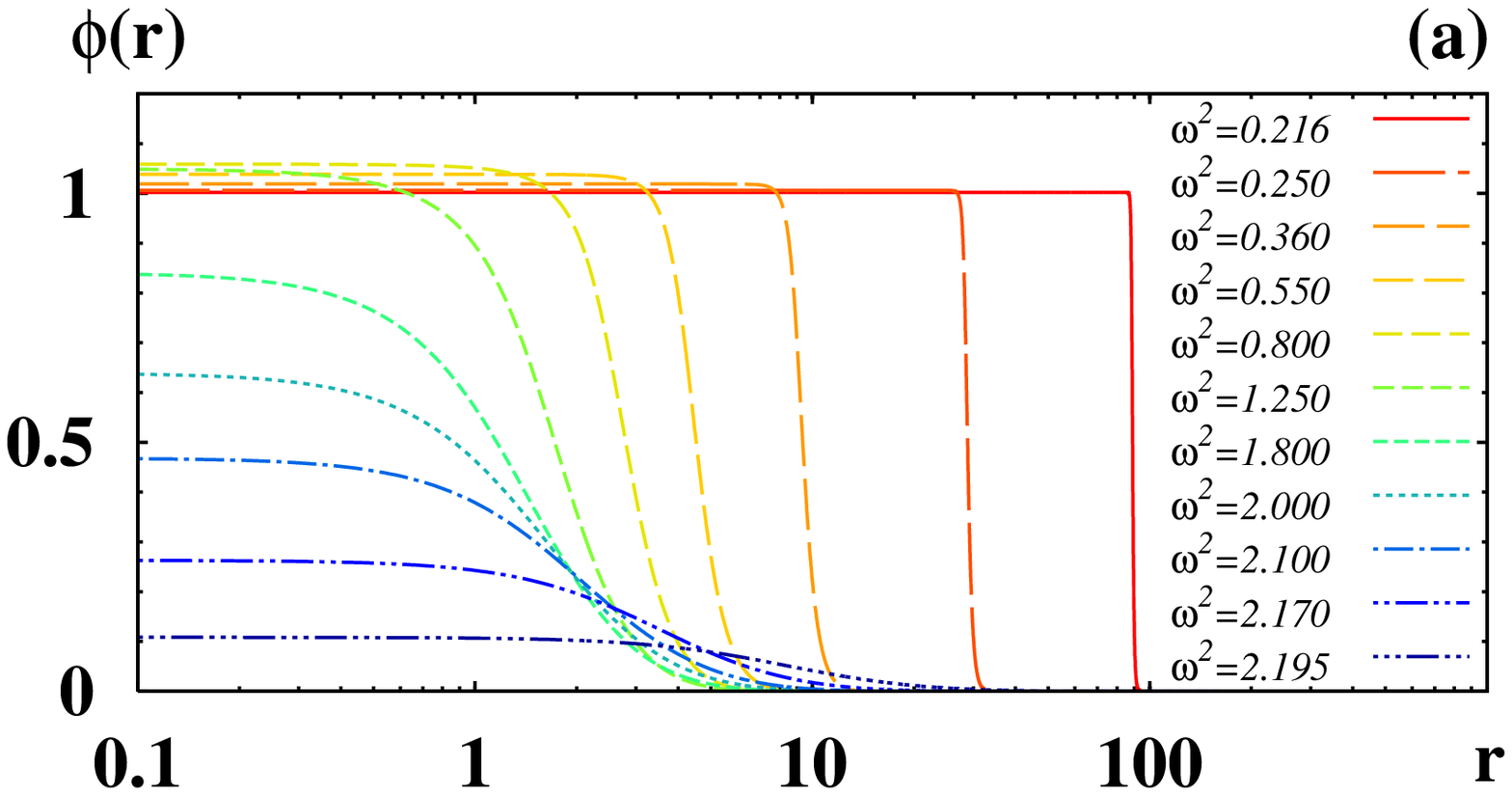}
\includegraphics[width=8.25cm]{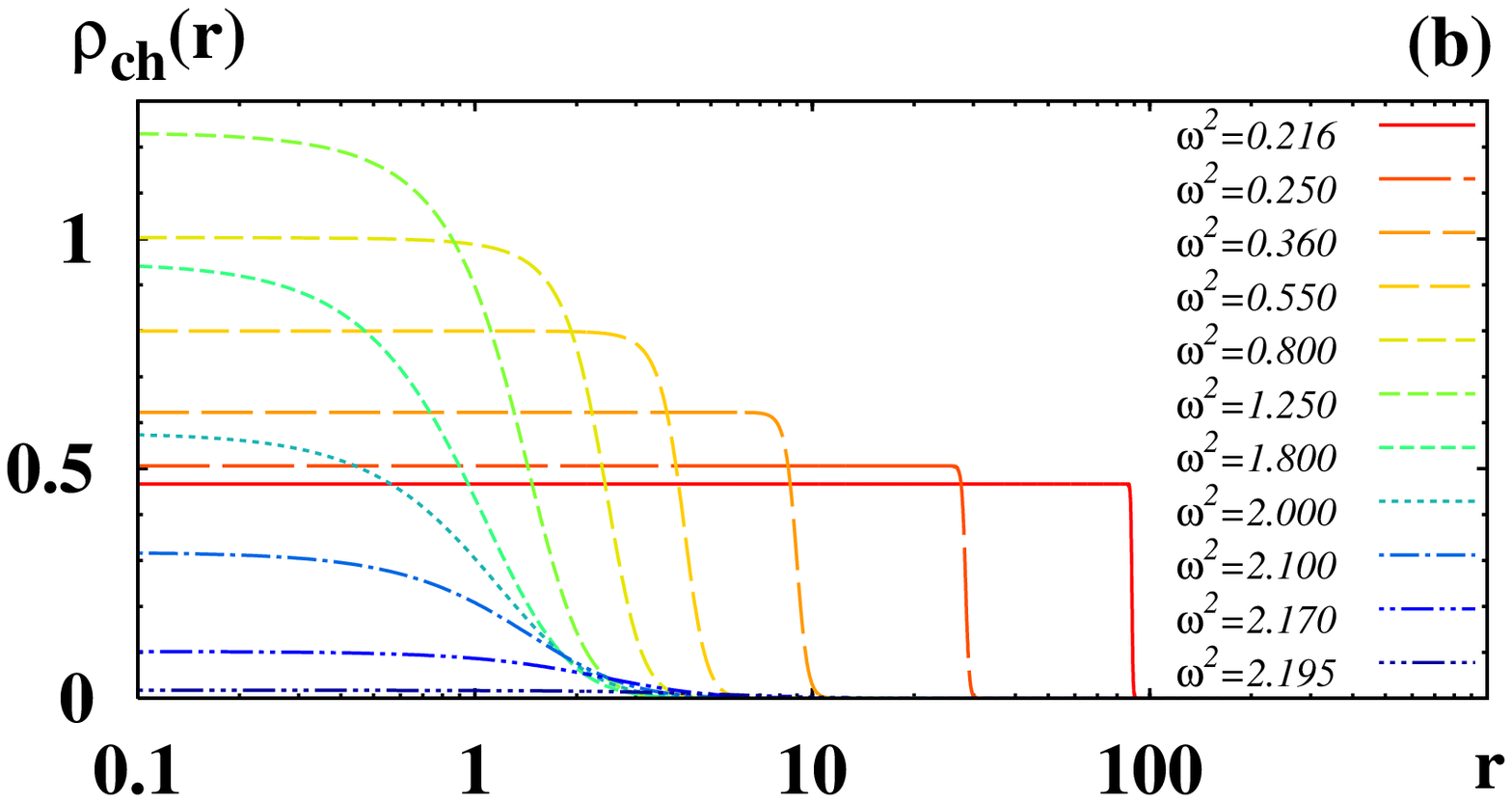}
\includegraphics[width=8.25cm]{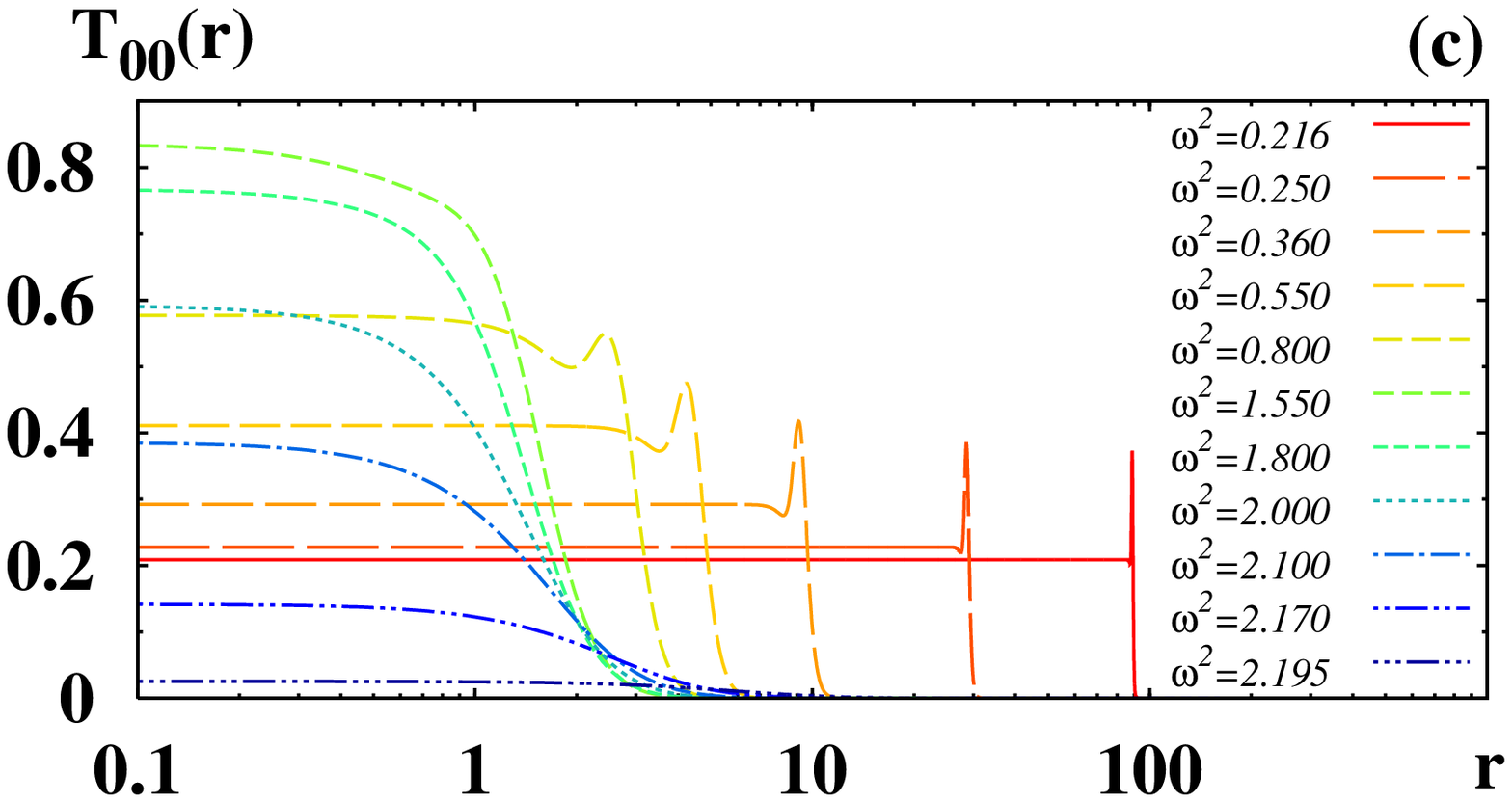}
\includegraphics[width=8.25cm]{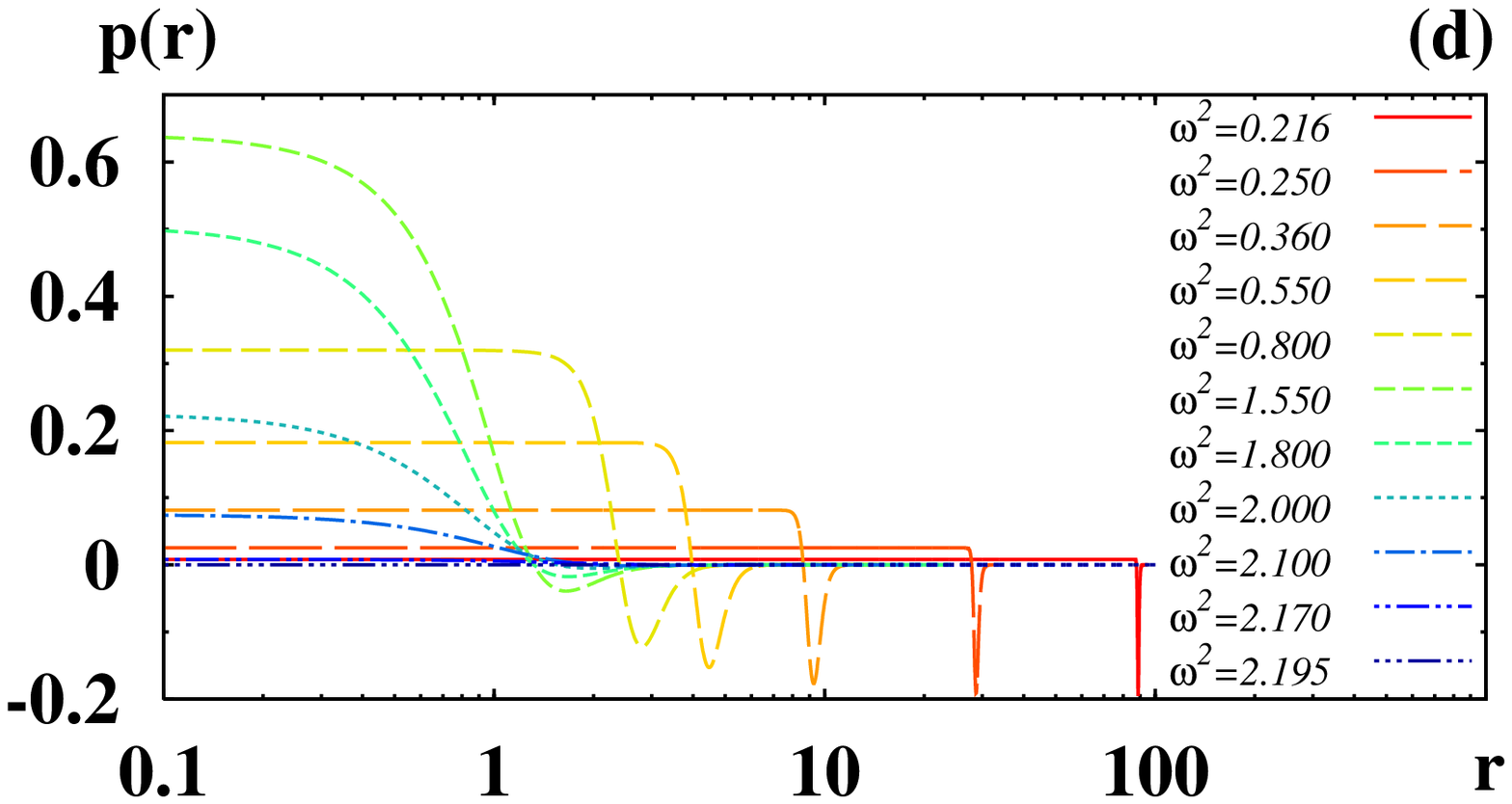}
\includegraphics[width=8.25cm]{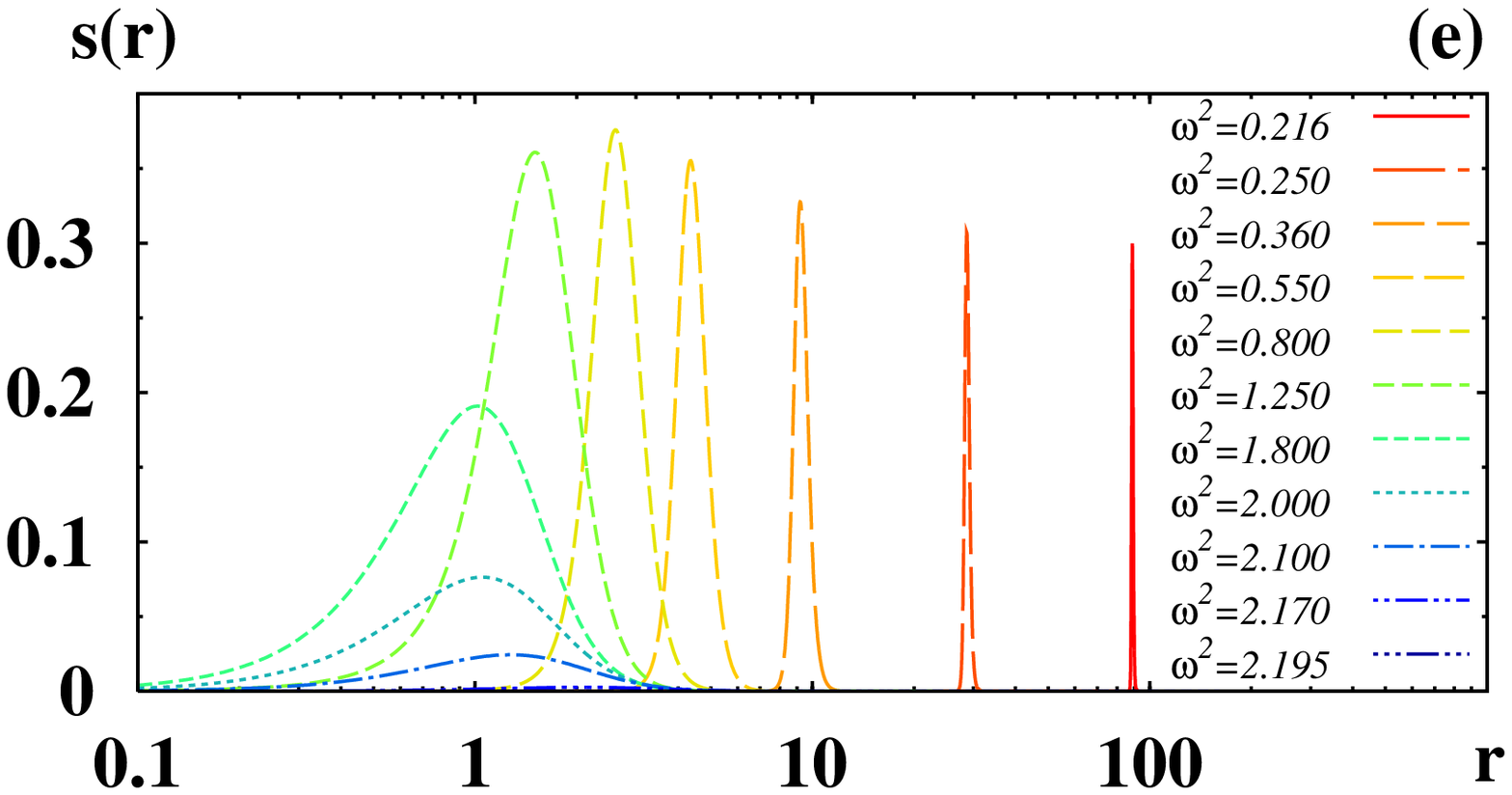}

\vspace{4mm}

\caption{\label{Fig-02:ground-state-densities}
Field $\phi(r)$, charge density $\rho_{\rm ch}(r)$, energy density
$T_{00}(r)$, pressure $p(r)$, shear force distribution $s(r)$
vs.\ $r$ for $Q$-ball ground state solutions for
selected values of $\omega$. 
}
\end{figure}

\subsection{\boldmath Solutions $\phi(r)$ and densities}
\label{Sec-3b:ground-state-densities}

In this section we describe the results for $\phi(r)$ and the 
various densities. Some of our observations concerning the behavior
of the densities in the limits $\omega\to\omega_{\rm min,max}$ will 
be made more rigorous in Secs.~\ref{Sec-4:limit-omega-to-min} and 
\ref{Sec-5:limit-omega-to-max}.

The ground state solutions $\phi(r)$, which are uniquely 
determined in terms of the initial values $\phi_0$ discussed 
in the previous section, are shown in
Fig.~\ref{Fig-02:ground-state-densities}a as functions of $r$
for selected values of $\omega$ in the range
$0.216\le\omega^2\le 2.195$ our numerics can handle.
We have chosen a logarithmic $r$-scale
to better show the features of all solutions in a single plot.
On a logarithmic scale the $\phi(r)$ are nearly constant for 
$r<0.1$, and have the small-$r$ behavior (\ref{Eq:asymp-small}).
Their large-$r$ asymptotics agrees with (\ref{Eq:asymp-large}).

With decreasing $\omega$ the solutions $\phi(r)$ remain nearly 
constant at their initial values in a region $0\le r < R_0$ and 
form increasingly long plateaus from which they then drop down to their
large-$r$ asymptotics (\ref{Eq:asymp-large}) over 
decreasingly narrow transition regions with thicknesses $\ll R_0$.
Here $R_0$ can be understood as the ``size'' of the $Q$-ball,
which  will be defined below more accurately. 
In the limit $\omega\to\omega_{\rm min}$ the field 
$\phi(r)\to\phi_0\,\Theta(R_0-r)$ where $R_0\to\infty$ and 
$\phi_0\to1$, see (\ref{Eq:phi-in-limiting-cases}) 
and App.~\ref{App-A:trivial-solutions}.
This behavior (``thin-wall limit'') can be strictly 
derived \cite{Coleman:1985ki}.
In the opposite limit, as $\omega$ increases, the solutions 
$\phi(r)$ become more wide-spread and their magnitude 
decreases, see (\ref{Eq:phi-in-limiting-cases}) and 
App.~\ref{App-A:trivial-solutions}.

Fig.~\ref{Fig-02:ground-state-densities}b shows the charge 
densities $\rho_{\rm ch}(r)=\omega\phi(r)^2$ as functions of $r$.
Also the charge densities exhibit for small $\omega$ extended
plateaus in the region $0\le r\lesssim R_0$ inside the $Q$-balls, 
and drop abruptly to zero outside.
For $\omega\to\omega_{\rm max}$ the charge densities become 
more wide-spread and their magnitudes show an
overall decrease.

Fig.~\ref{Fig-02:ground-state-densities}c shows the
energy densities $T_{00}(r)$, which look qualitatively 
similar to charge densities for $\omega\gtrsim 1$.
But for $\omega\lesssim 1$ the energy densities 
start to develop a ``bump'' around $R_0$, and as $\omega$ 
approaches $\omega_{\rm min}$ the bump becomes a characteristic ``spike.''  
The reason for that is that for $\omega\gtrsim 1$ the $Q$-balls are 
``diffuse'' objects, while for $\omega\lesssim 1$ they start to develop a
more and more well-defined ``edge.'' In fact, as $\omega\to\omega_{\rm min}$ 
the notions of a ``surface'' and ``surface tension'' become better
defined \cite{Coleman:1985ki}.
The characteristic bump/spike structure in $T_{00}(r)$ at
$r \sim R_0$ reflects the contribution of the surface energy. 
In the limit $\omega\to\omega_{\rm max}$ we
find $T_{00}(r)\to 0$. In the limit $\omega\to\omega_{\rm min}$ we
have $T_{00}(r)\to {\rm const}$ for $0\le r < R_0$ with
a surface energy contribution proportional to $\delta(r-R_0)$
with $R_0\to\infty$. 

Fig.~\ref{Fig-02:ground-state-densities}d shows the pressures $p(r)$ 
as functions of $r$. For all ground state solutions the pressures
are positive ``inside'' and ``negative'' outside in agreement with 
general expectations (\ref{Eq:pressure-small-large-r}).
We are now in the position to provide an exact definition of 
the scale $R_0$ as the point where the pressure vanishes, i.e.\
$p(R_0)=0$ with $0 < R_0 < \infty$.
The stability condition is fulfilled because the
integrals $\int_0^{R_0}\di r\,r^2 p(r)$ and
$\int_{R_0}^\infty\di r\,r^2 p(r)$ have opposite signs
and precisely cancel. Numerically the sum of these 2 
contributions normalized with respect to the sum of
their moduli is of ${\cal O}(10^{-6})$ or smaller.
In the limit $\omega\to\omega_{\rm max}$ we find $p(r)\to 0$.
For $\omega\to\omega_{\rm min}$ we obtain 
$p(r)\to {\rm const}$ for $0\le r < R_0$ with a surface energy 
contribution proportional to $-\,\delta(r-R_0)$
and a diverging $Q$-ball size $R_0$.

Fig.~\ref{Fig-02:ground-state-densities}e shows the shear 
forces $s(r)$ which are best suited to discuss
the concepts of ``diffuseness'' or ``edge''. From 
(\ref{Eq:asymp-small},~\ref{Eq:asymp-large}) we see 
$s(r)\to 0$ as $r\to 0$ or $r\to\infty$, and from 
Eq.~(\ref{Eq:shear}) we see it is an evidently positive
quantity, i.e.\ $s(r)$ must have a global maximum somewhere.
To determine the position of this maximum consider
$s^\prime(r)=2\phi^\prime(r)\phi^{\prime\prime}(r)$. 
Now due to the monotony property of $\phi(r)$ discussed
in Sec.~\ref{Sec-3a:Ueff}, we have $\phi^\prime(r)=0$ only 
at $r=0$ and at infinity. I.e.\ the maximum of $s(r)$ coincides
with the point where $\phi^{\prime\prime}(r)=0$. This change of curvature
occurs in the vicinity of the ``edge'' of the $Q$-ball, and
in the limit $\omega\to\omega_{\rm min}$ precisely at $r=R_0$ 
where $s(r)$ becomes proportional to a $\delta(r-R_0)$ with a 
coefficient related to the surface tension. As $\omega$ approaches 
$\omega_{\rm max}$ the shear force distribution becomes wider and wider,
which indicates a more and more diffuse ``edge'' of the $Q$-ball.

\begin{figure}[t]

\vspace{2mm}

  \includegraphics[width=8.6cm]{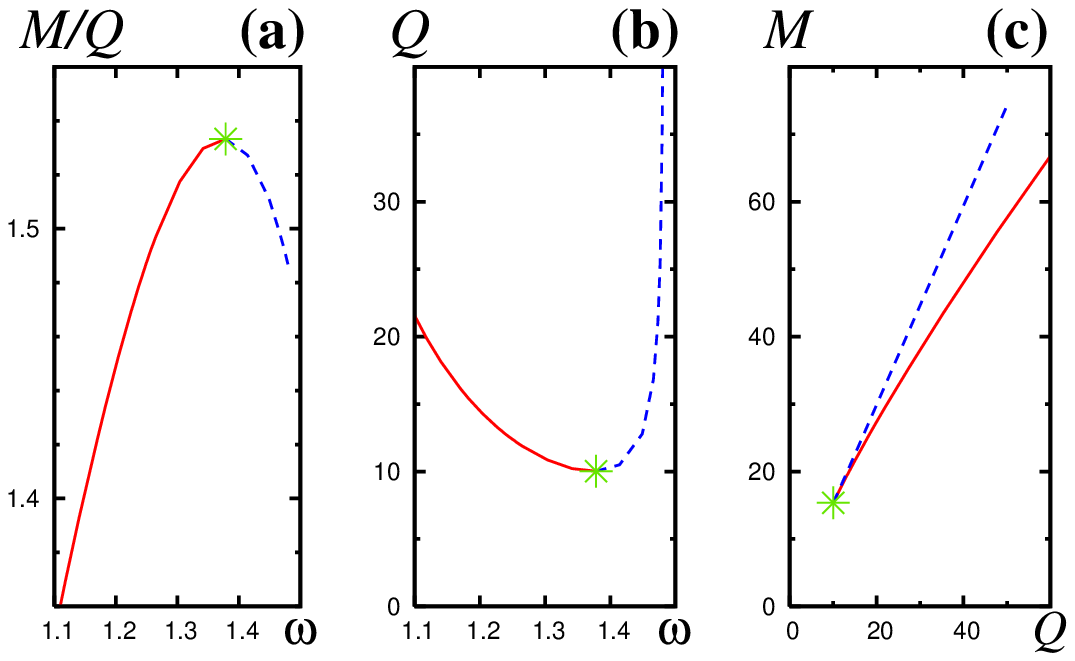}
  \caption{\label{Fig-03:omega-c}
    {\bf (a)} 
    $M/Q$ as function of $\omega$, with
    $(M/Q)^\prime(\omega) > 0$ in the range 
    $\omega < \omega_c \approx 1.38$ (solid line).
    {\bf (b)}
    $Q$ as function of $\omega$, with
    $Q^\prime(\omega) < 0$ for $\omega < \omega_c$ 
    (solid line).
    {\bf (c)}
    $M$ vs.\ $Q$, with $M^{\prime\prime}(Q)<0$ 
    in the branch denoted by the solid line.
    The solid (dotted) lines correspond to the region 
    of classically stable (unstable) $Q$-balls, see
    Eq.~(\ref{Eq:classical-stability}).}

\vspace{3mm}

  \includegraphics[width=8.6cm]{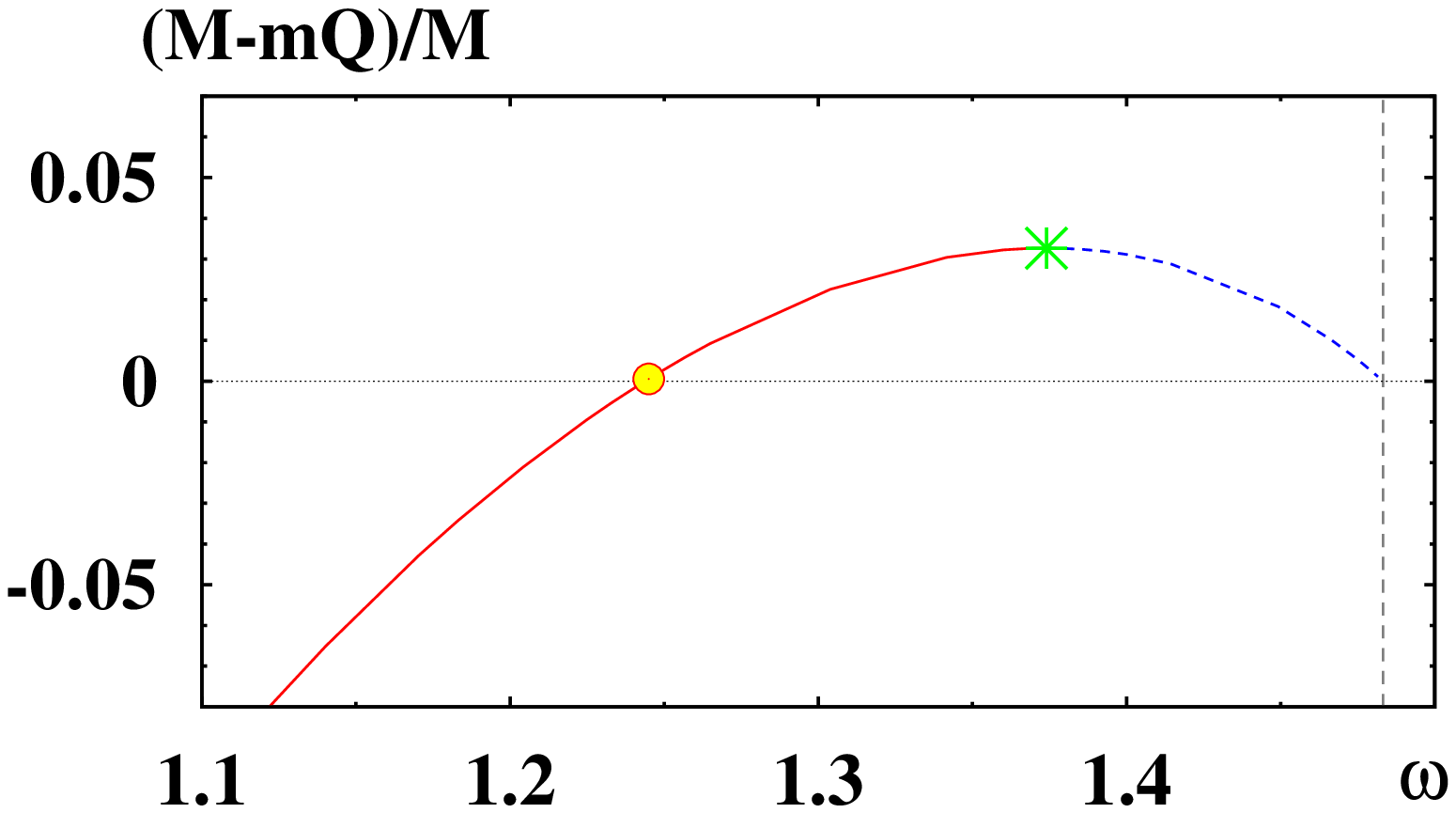}

  \caption{\label{Fig-04:omega-abs}
    $(M-m\,Q)/M$ as function of $\omega$ which exhibits a maximum at
    $\omega_c$. For $\omega<\omega_{\rm abs}$ we have $M-mQ<0$ and 
    the $Q$-balls are absolutely stable. 
    For $\omega_{\rm abs}<\omega<\omega_c$ the $Q$-balls are meta-stable,
    and for $\omega_c<\omega<\omega_{\rm max}$ they are unstable.
    $(M-m\,Q)$ approaches zero as $\omega\to\omega_{\rm max}$
    which is indicated by the vertical line.
}

\end{figure}

\subsection{\boldmath $\omega_c$ and $\omega_{\rm abs}$}
\label{Sec-3c:special-omega}

The frequencies $\omega_c$ and $\omega_{\rm abs}$ were 
discussed in the sequence of Eq.~(\ref{Eq:classical-stability}), 
and mentioned in the context of Fig.~\ref{Fig-01:eff-pot}.
In this Section we discuss how they appear in the numerical 
results. At the frequency $\omega=\omega_c\approx 1.38$ 
\begin{enumerate}
\item $(M/Q)(\omega)$ has a global maximum, 
      Fig.~\ref{Fig-03:omega-c}a,
\item $Q(\omega)$ has a global minimum, 
      Fig.~\ref{Fig-03:omega-c}b, 
\item $M(Q)$ has a branch point at $Q_c=Q(\omega_c)$,
      Fig.~\ref{Fig-03:omega-c}c.
\end{enumerate}
The frequency $\omega_c$ and charge $Q_c$ 
define classical stability. For $\omega<\omega_c$ we have 
$(M/Q)^\prime(\omega)>0$, and $Q^\prime(\omega)<0$.
For $Q>Q_c$ we have $M^{\prime\prime}(Q)<0$.
These are equivalent criteria  for the stability of $Q$-balls 
against small fluctuations. In Fig.~\ref{Fig-03:omega-c} the
branches of classically stable $Q$-balls are shown as 
solid lines, while the unstable branches are depicted as
dotted lines.

Classical stability is a necessary but not sufficient condition
for stability. For a $Q$-ball to be absolutely stable
it is required $M < mQ$, see Eq.~(\ref{Eq:absolute-stability}).
Fig.~\ref{Fig-04:omega-abs} shows $M-mQ$ normalized with
respect to $M$ as function of $\omega$. The quantity 
$(M-mQ)/M$ is negative for $\omega<\omega_{\rm abs}\approx1.245$ 
(and has a global maximum at $\omega=\omega_c$ which follows from 
the fact that $M(\omega)$ and $Q(\omega)$ have extrema there).
Thus, for  $\omega<\omega_{\rm abs}$ the Q-balls are absolutely
stable. 
In the region  $\omega_{\rm abs}<\omega<\omega_c$ they are 
meta-stable. For $\omega>\omega_c$ we have unstable $Q$-balls.

In the limit $\omega\to\omega_{\rm max}\equiv m$ 
one observes $M\to mQ$, see Fig.~\ref{Fig-04:omega-abs}.
This means the unstable $Q$-balls dissociate into $Q$-clouds,
i.e.\ into a gas of free quanta
\cite{Alford:1987vs}.

\begin{figure*}[ht!]
\includegraphics[width=17.5cm]{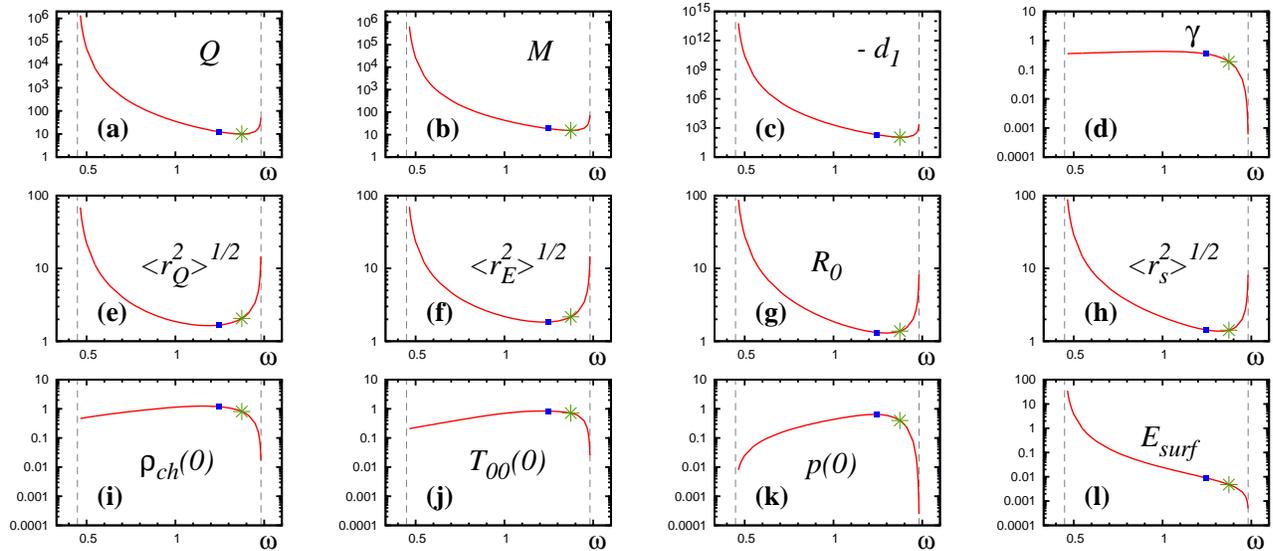}

\vspace{-4mm}

\caption{\label{Fig-05:ground-state-properties}
  Ground state properties of $Q$-balls as functions of $\omega$.
  (a) charge $Q$, Eq.~(\ref{Eq:charge}).
  (b) mass $M$, defined before Eq.~(\ref{Eq:T00}).
  (c) constant $d_1$, Eq.~(\ref{Eq:def-d1}).
  (d) ``surface tension'' $\gamma$, Eq.~(\ref{Eq:def-gamma-rs}).
  The mean square radii of 
  (e) charge and 
  (f) energy densities, Eq.~(\ref{Eq:def-mean-square-radii-E-M}), and
  (h) shear forces, Eq.~(\ref{Eq:def-gamma-rs}).
  (g) position $R_0$ of the zero of the pressure.
  The values of densities in the centers of $Q$-balls for
  (i) charge density, 
  (j) energy density,
  (k) pressure, Eqs.~(\ref{Eq:charge},~\ref{Eq:T00},~\ref{Eq:pressure}).
  (l) The ``surface energy'' defined in (\ref{Eq:virial-theorem-Ib}).
  The special value 
   $\omega_{\rm abs}\approx1.245$ 
  ($\omega_c\approx 1.38$) is marked by a square (star).
  The vertical lines indicate the limits
  $\omega_{\rm min}\approx 0.447$ and 
  $\omega_{\rm max}\approx 1.483$.}
\end{figure*}

\subsection{Ground state properties}
\label{Sec-3d:ground-state-properties}

In this Section we study ``global'' $Q$-ball properties, 
appropriate integrals 
of the ``local'' densities from  Sec.~\ref{Sec-3b:ground-state-densities}.
The numerical results are shown in Fig.~\ref{Fig-05:ground-state-properties}
which is organized as follows. The columns show as functions of $\omega$
quantities associated with (from left to right) the distributions of charge, 
energy, pressure and shear forces. 
Values of $\omega_{\rm abs}$, $\omega_c$, $\omega_{\rm min/max}$
are indicated in all plots.

Figs.~\ref{Fig-05:ground-state-properties}a--d show $Q$, $M$, $d_1$, $\gamma$. 
At $\omega=\omega_c$ charge $Q(\omega)$ and mass $M(\omega)$ exhibit 
global minima, see (\ref{Eq:classical-stability}), while in the 
vicinity of $\omega\approx\omega_c$ the ``surface tension'' $\gamma(\omega)$ 
exhibits the largest curvature, and $-d_1(\omega)$ a global minimum.

Figs.~\ref{Fig-05:ground-state-properties}e--h
show the different length scales of $Q$-balls:
square roots of the mean square radii of the 
charge and energy densities and shear forces, 
Eqs.~(\ref{Eq:def-gamma-rs},~\ref{Eq:def-mean-square-radii-E-M}),
and $R_0$ which is where $p(r)$ changes sign.
The behavior is qualitatively similar: all radii increase 
as $\omega\to\omega_{\rm min/max}$, and have global minima
around $\omega\approx\omega_{\rm abs}$
(for the minimum of $\la r_s^2\ra^{1/2}$ we cannot exclude that 
it is, within numerical accuracy, exactly at $\omega_{\rm abs}$).

Figs.~\ref{Fig-05:ground-state-properties}i--k
shows the charge density, energy density, and pressure
at the center of the $Q$-balls as functions of $\omega$.
These quantities exhibit maxima around $\omega_{\rm abs}$.

\begin{figure}[b!]
\includegraphics[width=8cm]{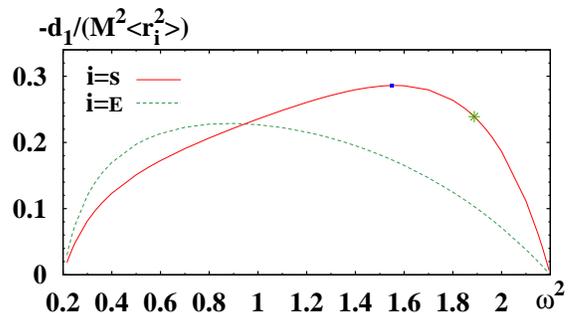}
\vspace{-4mm}
\caption{\label{Fig-06:d1-rescaled}
  $(-d_1)$ in units of $M^2\la r_i^2\ra$ ($i=s,E$) as function of $\omega^2$.
  These units take into account the true dimensionality of $d_1$.}
\end{figure}

Since $s(0)$ vanishes, it would make no sense to show this quantity in analogy 
with Figs.~\ref{Fig-05:ground-state-properties}i--k. Instead, in 
Fig.~\ref{Fig-05:ground-state-properties}l we show the ``surface energy,'' 
Eq.~(\ref{Eq:virial-theorem-Ib}), as function of $\omega$. 
As $\omega$ increases from $\omega_{\rm min}$ to 
$\omega_{\rm max}$ the surface energy decreases monotonically,
which is compatible with the view that with increasing
$\omega$ the $Q$-ball becomes a more and more diffuse
object, see Sec.~\ref{Sec-3b:ground-state-densities},
such that the role of a ``surface energy'' become less
and less important.
$E_{\rm surf}(\omega)$ changes the curvature
at the point $\omega=\omega_c$ within numerical accuracy.

Some of the quantities vary strongly with $\omega$, for instance  
$d_1$ extends over 12 orders of magnitude. This is not surprising 
since we compare $Q$-balls with different masses and sizes. 
For each quantity one could find ``natural units'' in order 
to make (from this point of view) the comparisons quantitatively 
more meaningful. The dimensionless constant $d_1$ has the natural units 
(mass$\times$length)$^2$. To see this notice that it can be obtained from 
$d_1=\frac13M\int\di^3x\,r^2s(r)$, and $\int\di^3x\,s(r)$ has dimension mass, 
since $s(r)$ and $T_{00}(r)$ have the same dimensions. Thus, one choice of 
``natural units'' to measure $d_1$ could be $M^2\la r_i^2\ra$ with $i=s,E$. 
In Fig.~\ref{Fig-06:d1-rescaled} we see that in these units the constant 
$d_1$ varies much more moderately. In fact, for $Q$-balls of all $\omega$ 
we find that $0 < (-d_1)/(M^2\la r_s^2\ra) < 0.3$ holds. 

\newpage
\section{\boldmath The limit $\omega\to\omega_{\rm min}$ }
\label{Sec-4:limit-omega-to-min}

\begin{figure}[b!]
\includegraphics[width=6cm]{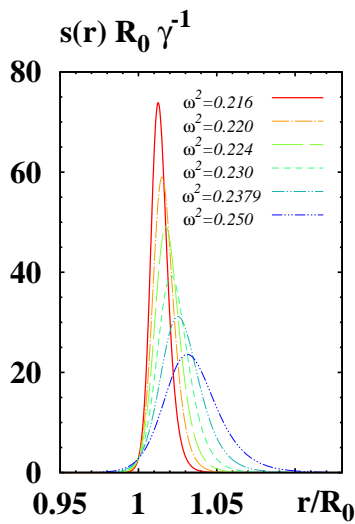}\\

\vspace{-4mm}

\caption{\label{Fig-07:s-r-liqid-drop-lim}
  $s(r)R_0/\gamma$ as function of $r/R_0$ in the ``edge region,''  for 
  selected values of $\omega$ in the range $0.216\le \omega^2 \le 0.25$. 
  The curves are scaled
  such that the areas under the graphs are normalized to unity.
  The figure shows that with $\omega^2$ approaching $\omega_{\rm min}^2=0.2$
  the shear forces approach their liquid drop limit $s(r)=\gamma\,\delta(r-R_0)$
  where $\gamma$ denotes the surface tension, and $R_0$ the position
  at which the pressure vanishes.}
\end{figure}

In this section we discuss $Q$-ball properties in 
the limit $\omega\to\omega_{\rm min}$, 
the so-called  ``thin wall'' limit.
In this limit the solutions describe objects of increasing size $R$ 
with uniform charge distribution for $r<R$, which drops to zero over 
a narrow transition region (``thin wall'') \cite{Coleman:1985ki}.
In some sense the $Q$-balls resemble liquid drops. 

\begin{figure}[b!]
\includegraphics[width=6.3cm]{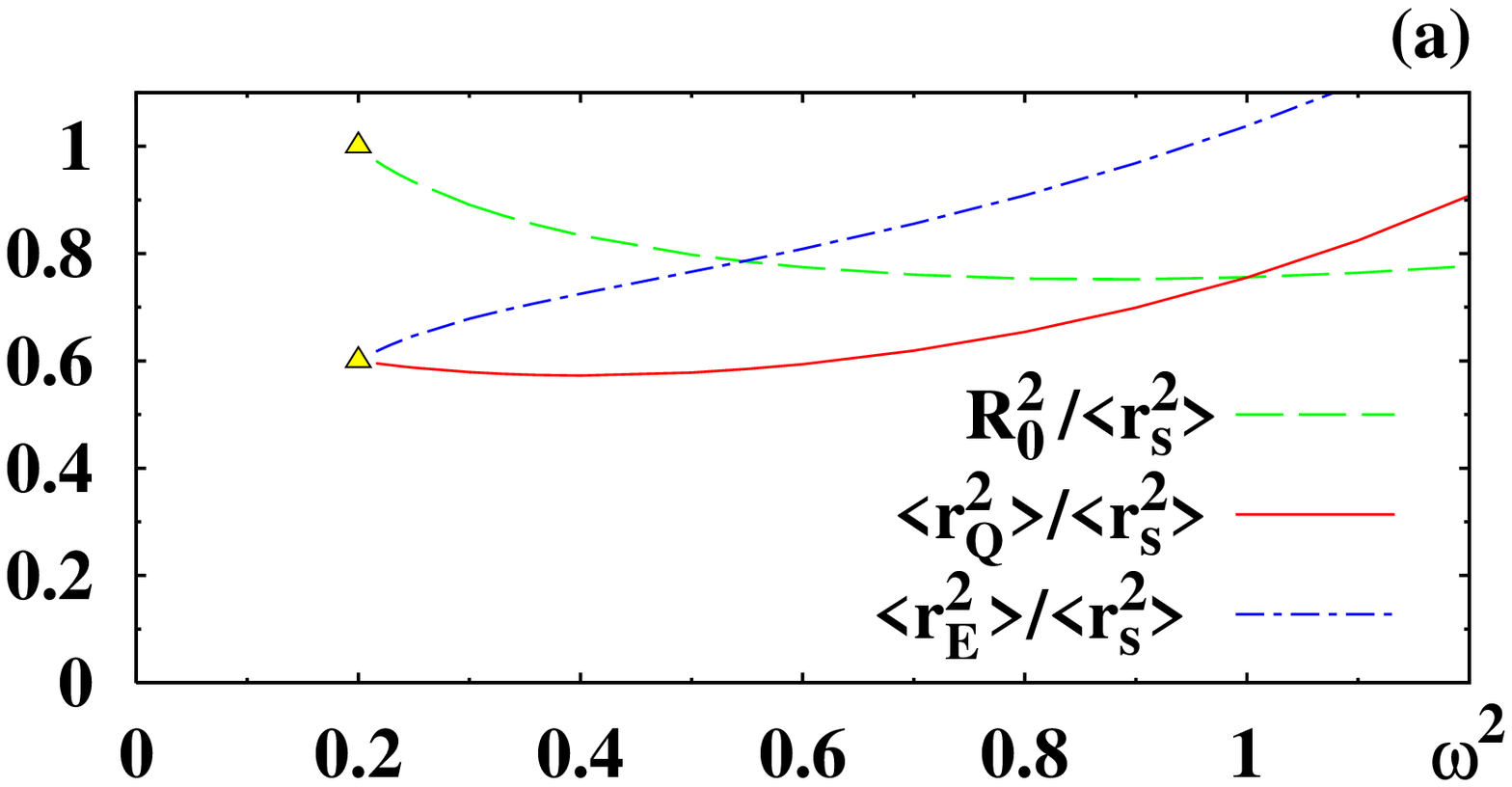}\\
\includegraphics[width=6.2cm]{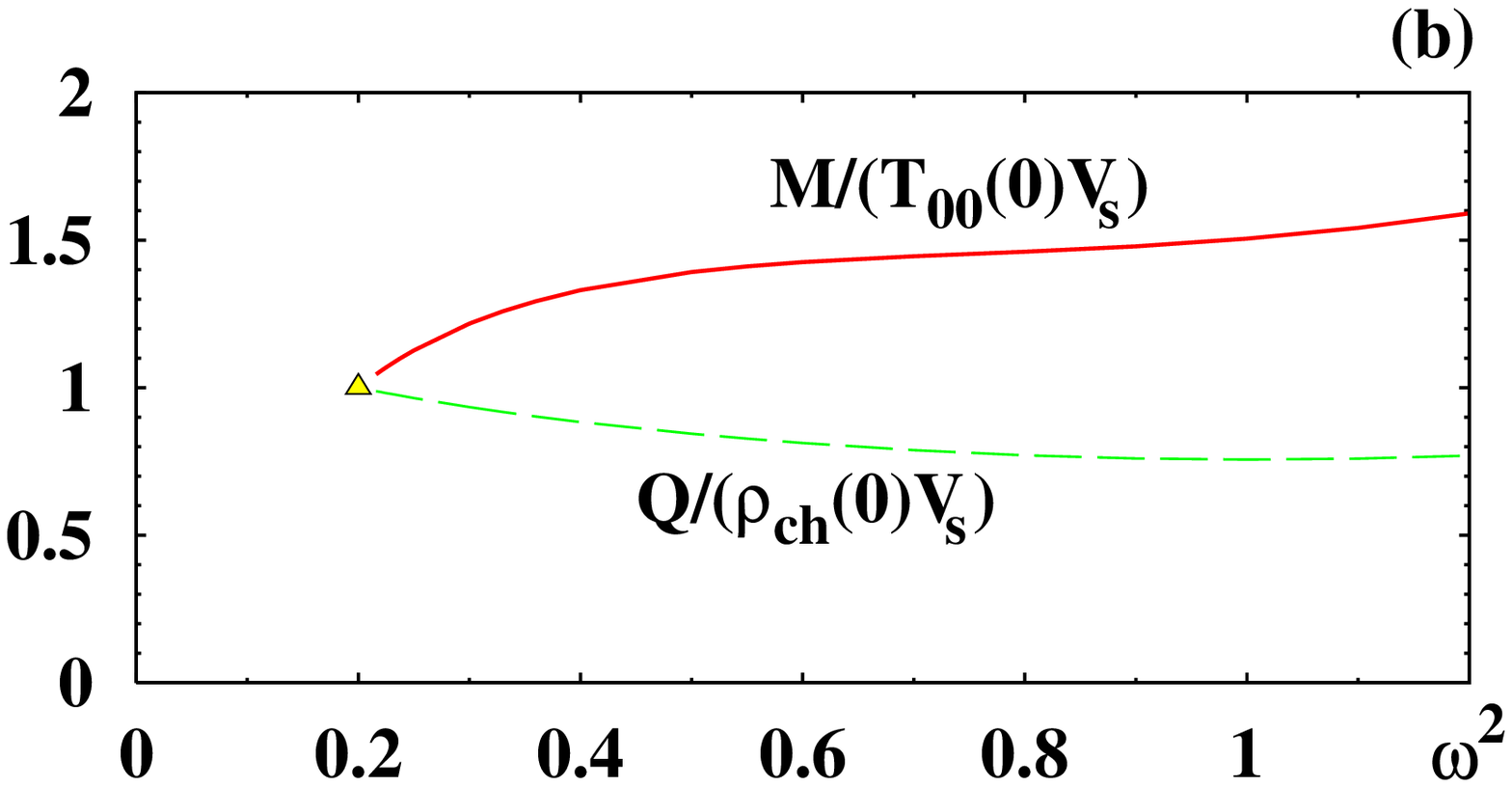}\\
\includegraphics[width=6.2cm]{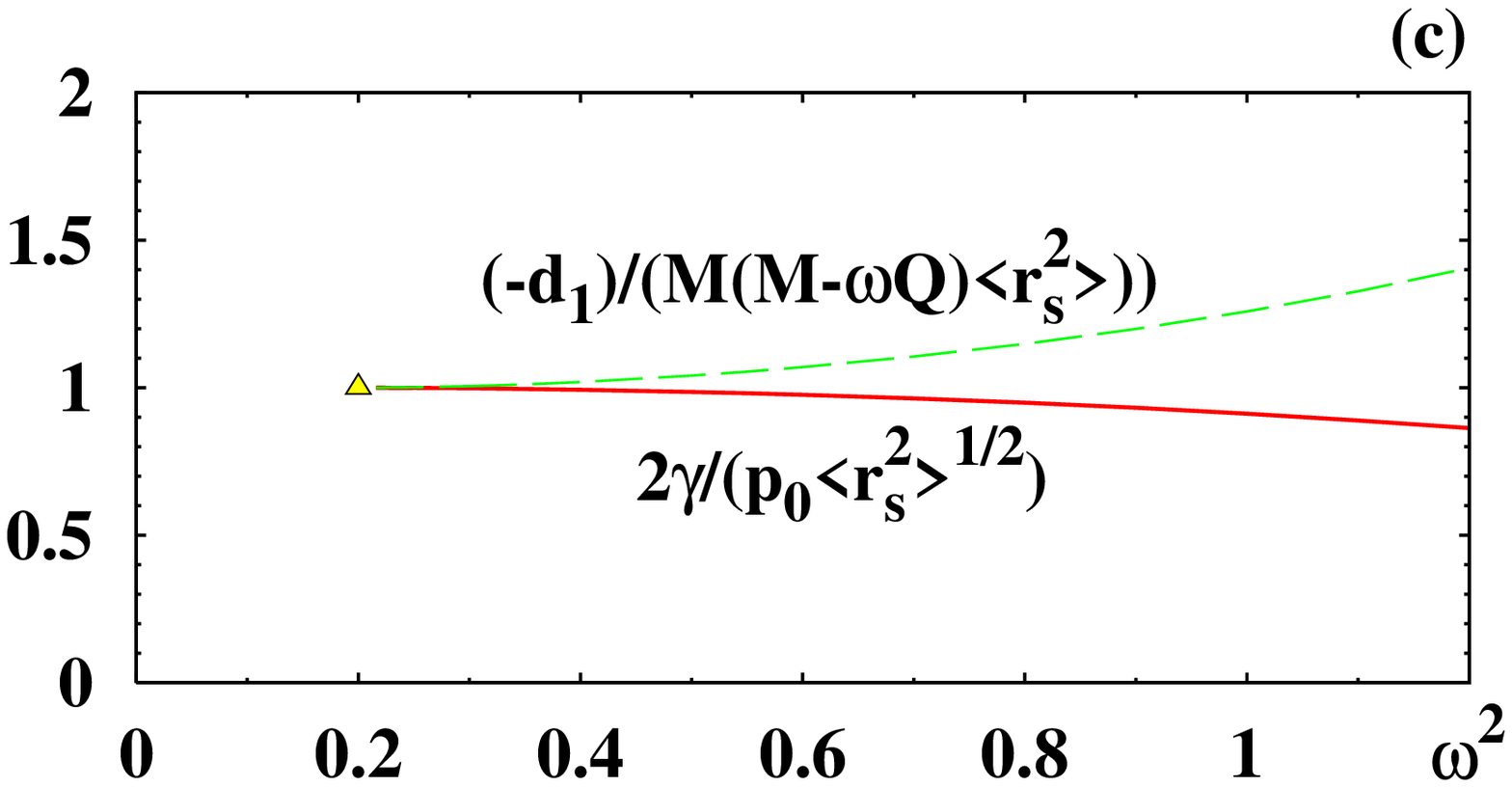}\\
\includegraphics[width=6.2cm]{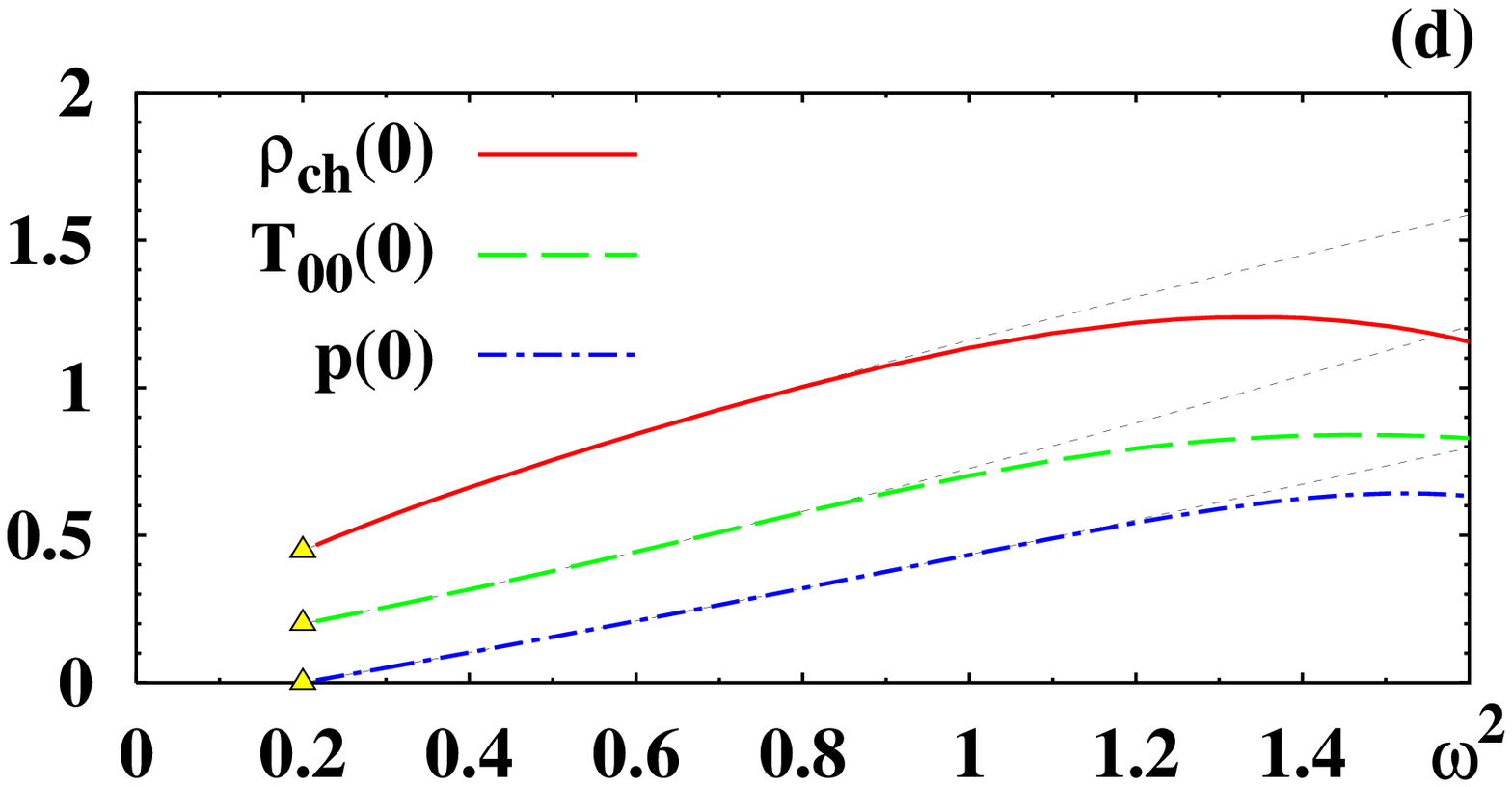}

\caption{\label{Fig-08:liq-drop}
  $Q$-ball properties as functions of $\omega^2$
  in the limit $\omega\to\omega_{\rm min}$.
  (a) the ratios $R_0^2/\la r_s^2\ra$, $\la r_Q^2\ra/\la r_s^2\ra$,  
  $\la r_E^2\ra/\la r_s^2\ra$.
  (b) $M/(T_{00}(0)V_s)$ and $Q/(\rho_{\rm ch}(0)V_s)$.
  (c) $2\gamma/(p(0)\la r_s^2\ra^{1/2})$ and
  $(-\,d_1)/(M(M-\omega Q)\la r_s^2\ra)$.
  (d) $\rho_{\rm ch}(0)$, $T_{00}(0)$, $p(0)$.
  The solid, dashed, dashed-dotted lines show our 
  numerical results, which approach the predicted 
  limits marked by symbols.
  In Fig.~\ref{Fig-08:liq-drop}d the thin lines are
  the analytic results derived from 
  Eq.~(\ref{Eq:constant-solutions}).}


\end{figure}

In a liquid drop of the size $R$ the pressure distribution
is $p(r)=p_0\,\Theta(R-r)-\frac13\,p_0 R \delta(R-r)$ where 
$p_0$ denotes the constant pressure inside the drop.
This $p(r)$ satisfies the stability condition
(\ref{Eq:stability-condition}).
The shear forces are given by $s(r)=\gamma\,\delta(R-r)$,
and the differential equation (\ref{Eq:diff-eq-s-p})
leads to the Kelvin relation $\gamma=\frac12\,p_0 R$. 

In the following we will ``test'' the predictions from the
liquid drop picture using our numerical results, and derive 
analytically relations valid in the limit
$\omega\to\omega_{\rm min}$.

In Fig.~\ref{Fig-02:ground-state-densities} we have seen that the
solutions and densities approach the expected liquid drop shapes,  
see Sec.~\ref{Sec-3b:ground-state-densities}. Let us highlight
here the shear force distribution.
Fig.~\ref{Fig-07:s-r-liqid-drop-lim} shows $s(r)R_0/\gamma$ as function 
of $r/R_0$ in the ``edge region'' for selected $\omega$ close to 
$\omega_{\rm min}$.
The curves are scaled such that the areas under the graphs are 
normalized to unity. Clearly, as $\omega$ approaches $\omega_{\rm min}$
the shear force distributions peak more and more strongly in a narrow
region concentrated around $r/R_0\approx 1$. The ``edge region'' makes
makes $5\,\%$ and less of the size of the $Q$-ball, as expected.

That the $Q$-ball size diverges as $\omega\to\omega_{\rm min}$
is apparent from Figs.~\ref{Fig-05:ground-state-properties}e--h.
Our first quantitative expectation is that both the radius $R_0$ 
describing the position of the zero of the pressure and 
$\la r_s^2\ra^{1/2}$ characterize equally well the position 
of the ``edge'' of the $Q$-ball. Hence we expect these radii
to coincide for $\omega\to\omega_{\rm min}$, i.e.\
\be\label{Eq:prediction-liquid-drop-R0-rs}
      \lim\limits_{\omega\to\omega_{\rm min}}
      \frac{\la r_s^2\ra}{R_0^2} = 1\,.
\ee
The numerical results in Fig.~\ref{Fig-08:liq-drop}a 
support Eq.~(\ref{Eq:prediction-liquid-drop-R0-rs}).

In the following we choose $\la r_s^2\ra^{1/2}$ as a reference 
length scale for the size of the $Q$-ball in the liquid drop limit, 
and define the ``surface'' and ``volume'' of a $Q$-ball as 
\be\label{Eq:def-area-volume}
       A_s = 4\,\pi\, \la r_s^2\ra\, , \;\;\;
       V_s = \frac{4\,\pi}{3}\, \la r_s^2\ra^{3/2} \, .
\ee
The charge distribution becomes $\rho_{\rm ch}(r) = \rho_0\Theta(R-r)$
in the liquid drop limit, yielding $\la r_{\rm ch}^2\ra=\frac35 R^2$ for 
the mean square charge radius in Eq.~(\ref{Eq:def-mean-square-radii-E-M}). 
The situation is analog for $\la r_E^2\ra$, although $T_{00}(r)$ 
has a $\delta$-function-type ``spike''
at $r=R$ due to surface energy, as can
be seen in Fig.~\ref{Fig-02:ground-state-densities}c.
But the surface energy is proportional to $R^2$,
while the contribution of the constant bulk matter 
density inside the drop is proportional to $R^3$, so 
the influence of the spike can be neglected for a large drop.
Hence, we expect 
\be\label{Eq:prediction-liquid-drop-rch-rE}
      \lim\limits_{\omega\to\omega_{\rm min}}
      \frac{\la r_{\rm ch}^2\ra}{\la r_s^2\ra} = \frac35\,,\;\;\;
      \lim\limits_{\omega\to\omega_{\rm min}}
      \frac{\la r_E^2\ra}{\la r_s^2\ra} = \frac35\,,
\ee
and the numerical results in Fig.~\ref{Fig-08:liq-drop}a
support this. For the above discussed reasons we can furthermore 
expect 
\ba
      \lim\limits_{\omega\to\omega_{\rm min}}
      \frac{M}{T_{00}(0)\, V_s} &=& 1 \,,\nonumber\\
      \lim\limits_{\omega\to\omega_{\rm min}}
      \frac{Q}{\rho_{\rm ch}(0\, )V_s} &=& 1 \,.
      \label{Eq:prediction-liquid-mass-charge}
\ea
which is also confirmed, see Fig.~\ref{Fig-08:liq-drop}b.

Surface and surface tension $\gamma$ are abstract 
notions for arbitrary $Q$-balls which are {\sl defined} 
through Eq.~(\ref{Eq:def-gamma-rs}). One way to check the 
usefulness of those definitions provides the Kelvin relation, which
implies the expectation
\be\label{Eq:prediction-liquid-drop-I}
      \lim\limits_{\omega\to\omega_{\rm min}}
      \frac{2\gamma}{p_0\la r_s^2\ra^{1/2}} = 1\,.
\ee
Notice that because of our definitions
(\ref{Eq:def-gamma-rs},~\ref{Eq:def-area-volume})
and (\ref{Eq:M-omegaQ-Esurf}), we always have the
relation
\be\label{Eq:helpful-relation}
      \frac{2\gamma}{p_0\la r_s^2\ra^{1/2}} = 
      \frac{2(M-\omega \,Q)}{p_0\,V_s} \,.
\ee
From (\ref{Eq:def-d1-pressure}) we obtain
for a liquid drop $d_1^{\rm drop}=-\frac{4\pi}{3}M\gamma R^4$.
Inserting here the expression for  $\gamma$ 
from (\ref{Eq:helpful-relation}) yields
\be\label{Eq:prediction-liquid-drop-d1-I}
      \lim\limits_{\omega\to\omega_{\rm min}}
      \frac{(-1)d_1}{M(M-\omega \,Q) \la r_s^2\ra} = 1\;.
\ee
Fig.~\ref{Fig-08:liq-drop}b confirms both relations
(\ref{Eq:prediction-liquid-drop-I}) and 
(\ref{Eq:prediction-liquid-drop-d1-I}).

Next let us focus on the center properties of $Q$-balls. 
For $\omega\to\omega_{\rm min}$ the limiting value of 
the field $\phi(r)$ at $r=0$ assumes the value 
$\phi_{\rm const}^2=B/(2C)$, 
see (\ref{Eq:phi-in-limiting-cases}) and 
App.~\ref{App-A:trivial-solutions}.
For our potential and choice of parameters this means
\ba
   &&   \lim\limits_{\omega\to\omega_{\rm min}} \rho_{\rm ch}(0) 
      = \omega_{\rm min}\,\phi_{\rm const}^2=\sqrt{0.2},\;\;\nonumber\\
   &&   \lim\limits_{\omega\to\omega_{\rm min}} T_{00}(0) 
      = \omega_{\rm min}^2\,\phi_{\rm const}^2 =0.2\,,\;\; \nonumber\\
   &&   \lim\limits_{\omega\to\omega_{\rm min}} \;\,p(0)\;\: 
      = 0\,,
      \label{Eq:prediction-liquid-drop-densities} 
\ea
which is supported by our results in Fig.~\ref{Fig-08:liq-drop}d. 
The result for $p(0)$ is derived alternatively in 
App.~\ref{App-B:bound-on-pressure}.

\begin{figure}[b!]

\includegraphics[width=5.6cm]{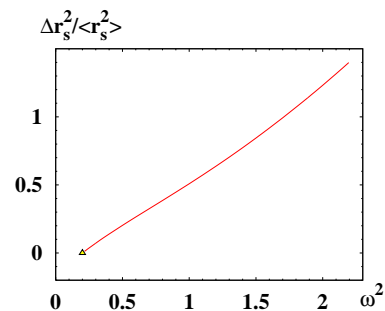}

\caption{\label{Fig-09:liq-drop-width-of-wall}
  The ratio $\Delta r_s^2/\la r_s^2\ra$ characterizing
  the relative size of the ``wall thickness,'' 
  as function of $\omega^2$ (solid line). In the limit 
  $\omega\to\omega_{\rm min}$ (``thin-wall limit'')
  $\Delta r_s^2/\la r_s^2\ra\to 0$ (marked by the symbol).
  The numerical results support this expectation.}

\end{figure}

We can go a step further and derive predictions from the liquid drop 
picture for the densities in (\ref{Eq:prediction-liquid-drop-densities})
also for $\omega\neq\omega_{\rm min}$. 
This can be done because the probably most important features for
$\omega > \omega_{\rm min}$ are the finite size of the
$Q$-ball, and its diffuse ``edge.'' 
But these features become important ``far away'' from the
$Q$-ball center. So one would expect the liquid drop
approach to give useful approximations for 
$\rho_{\rm ch}(0)$, $T_{00}(0)$, and $p(0)$ not only
for $\omega=\omega_{\rm min}$ but also for $\omega$ 
in some vicinity of $\omega_{\rm min}$.

The result for $\phi_{\rm const}(\omega)$ follows from 
Eq.~(\ref{Eq-App:triv-sol-II}) using the plus sign, and
can be written as 
\be\label{Eq:constant-solutions}
      \phi_{\rm const}^2(\omega) = \frac{B}{C}\,
      \biggl(\frac13+\frac16\,
      \sqrt{1+\frac{6C}{B^2}(\omega^2-\omega_{\rm min}^2)}\,\biggr)\;.
\ee
Fig.~\ref{Fig-08:liq-drop}d shows that (\ref{Eq:constant-solutions}) 
provides excellent approximations for the exact $\omega$-dependencies 
of $\rho_{\rm ch}(0)$, $T_{00}(0)$, $p(0)$ from up to $\omega^2\lesssim 1$.
The reason for that can be seen in Fig.~\ref{Fig-01:eff-pot}.
In our potential, the ``particles'' (in the ``particle motion'' 
picture of Sec.~\ref{Sec-3a:Ueff}) are released very close 
to the respective maxima of $U_{\rm eff}$, and this is what
Eq.~(\ref{Eq:constant-solutions}) actually describes,
see App.~\ref{App-A:trivial-solutions}.

This brings us to another test of the liquid drop picture:
as $\omega\to\omega_{\rm min}$ we expect the ``edge'' 
of the $Q$-ball to become more and more well-defined. 
We have seen this in Fig.~\ref{Fig-07:s-r-liqid-drop-lim}, but
this observation can be made quantitative by defining the
thickness of the edge region as
\be\label{Eq:thickness-of-wall}
     (\Delta r^2_s)^2 = \la\la (r^2-\la\la r^2\ra\ra)^2\ra\ra
                      = \la\la  r^4 \ra\ra - \la\la r^2\ra\ra^2 \ge 0
\ee
where we introduced, c.f.\ App.~\ref{App-C:alternative-expressions-gamma-rs2} 
\be\label{Eq:def-mean-rs-radii}
\la\la r^n\ra\ra=\frac{\int_0^\infty\di r\;r^n s(r)}{\int_0^\infty\di r\;s(r)}
\;.
\ee
With this definition we can formulate the expectation that in the
``thin wall limit'' the relative size of the ``edge region'' vanishes  
\be
      \lim\limits_{\omega\to\omega_{\rm min}}
      \frac{\Delta r^2_s}{\la r^2_s\ra} \to 0\,.
\ee
This is supported by the numerical results, see
Fig.~\ref{Fig-09:liq-drop-width-of-wall}.

Finally we turn our attention to the behavior of integrated quantities 
for $\omega\to\omega_{\rm min}$. In this limit $Q$, $M$, $d_1$ 
and the radii diverge, as shown in Fig.~\ref{Fig-05:ground-state-properties},
although certain ratios of these quantities remain finite, and follow 
the predictions from the liquid drop picture, see the discussion above.
The key to understand quantitatively the behavior of 
these quantities in the limit  $\omega\to\omega_{\rm min}$
is the surface tension $\gamma$, the only ``integrated''
$Q$-ball property which remains finite in this limit.

\begin{figure}[t!]

\vspace{-3mm}

\includegraphics[width=6.5cm]{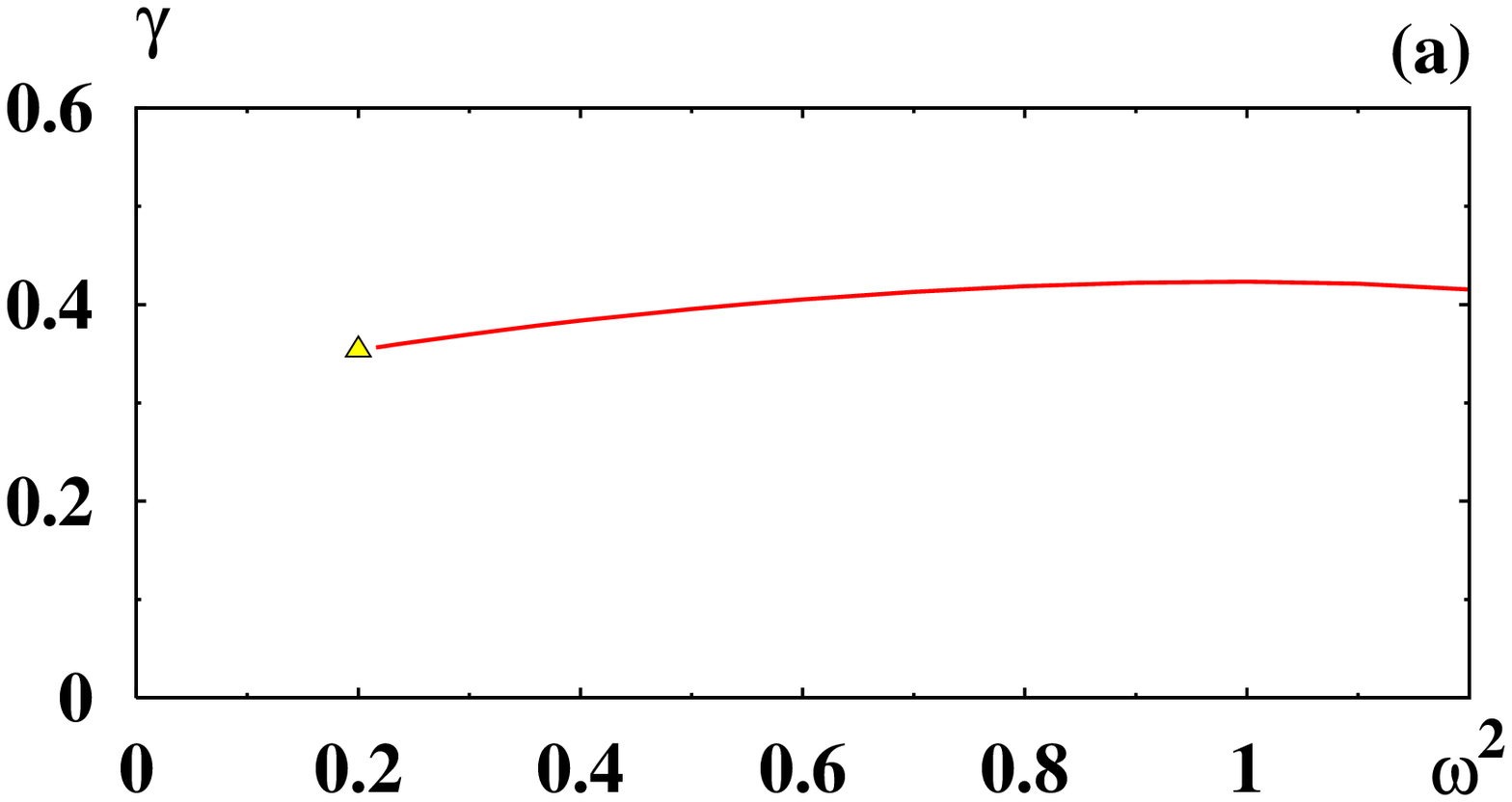}\\
\includegraphics[width=6.5cm]{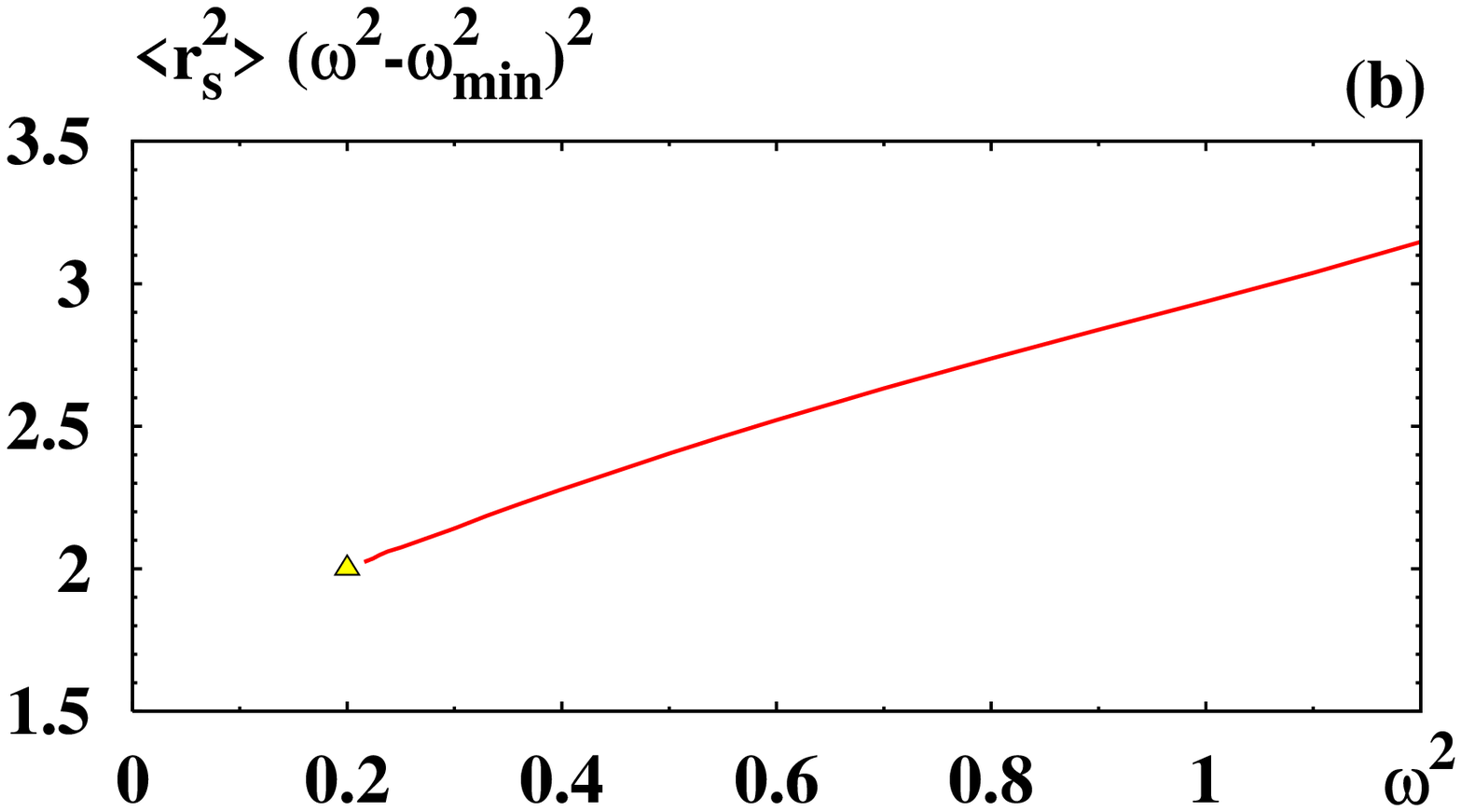}\\
\includegraphics[width=6.5cm]{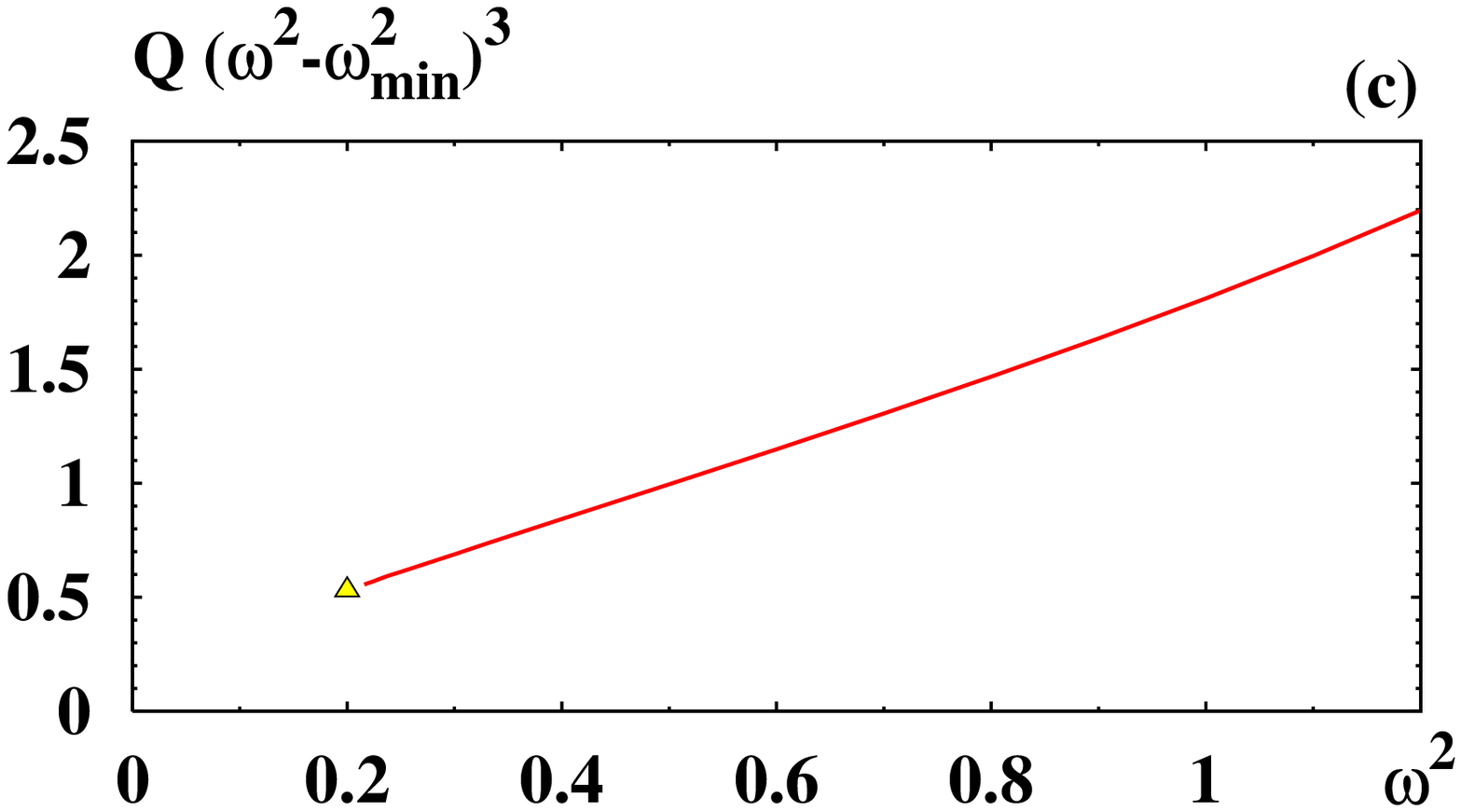}\\
\includegraphics[width=6.5cm]{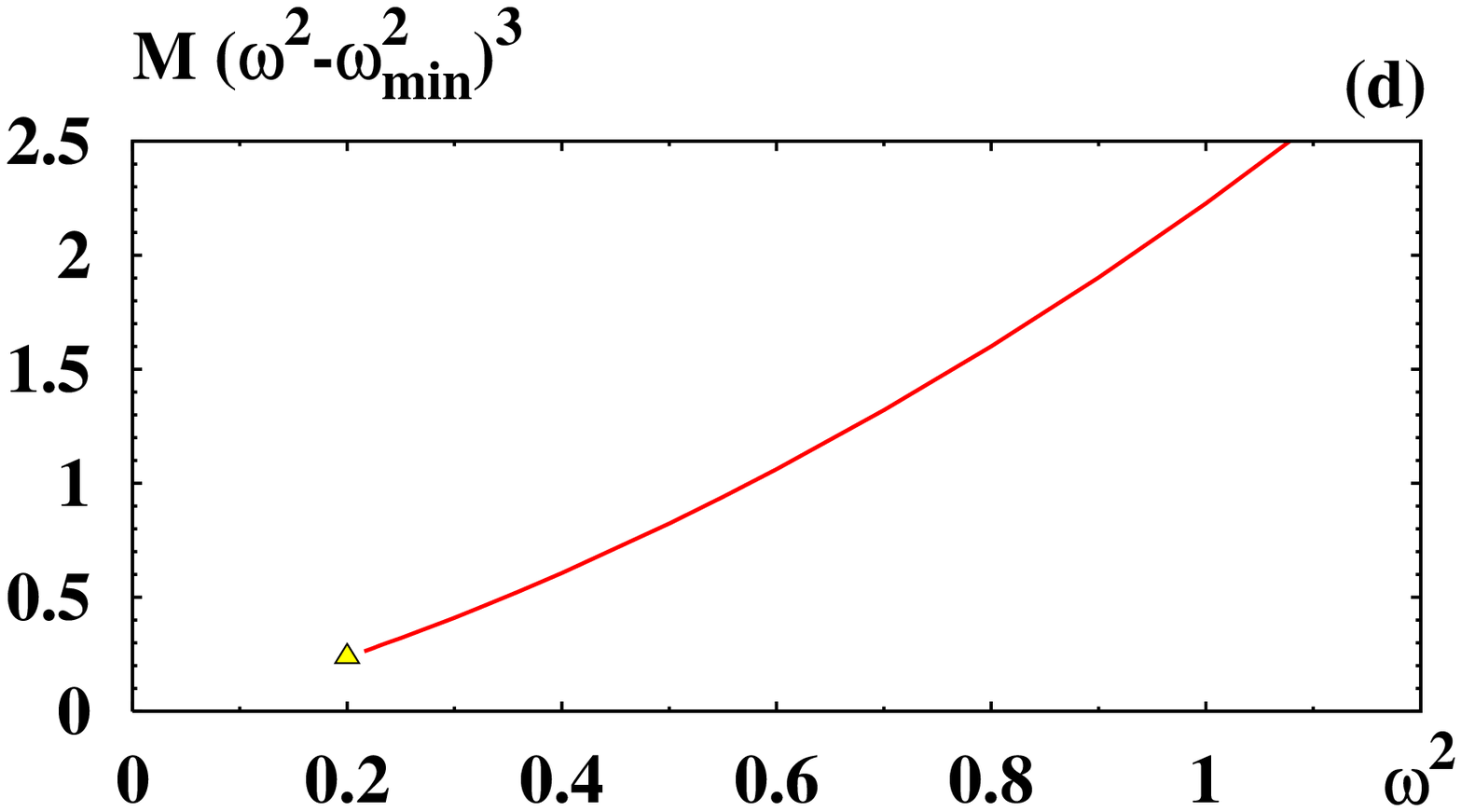}\\
\includegraphics[width=6.5cm]{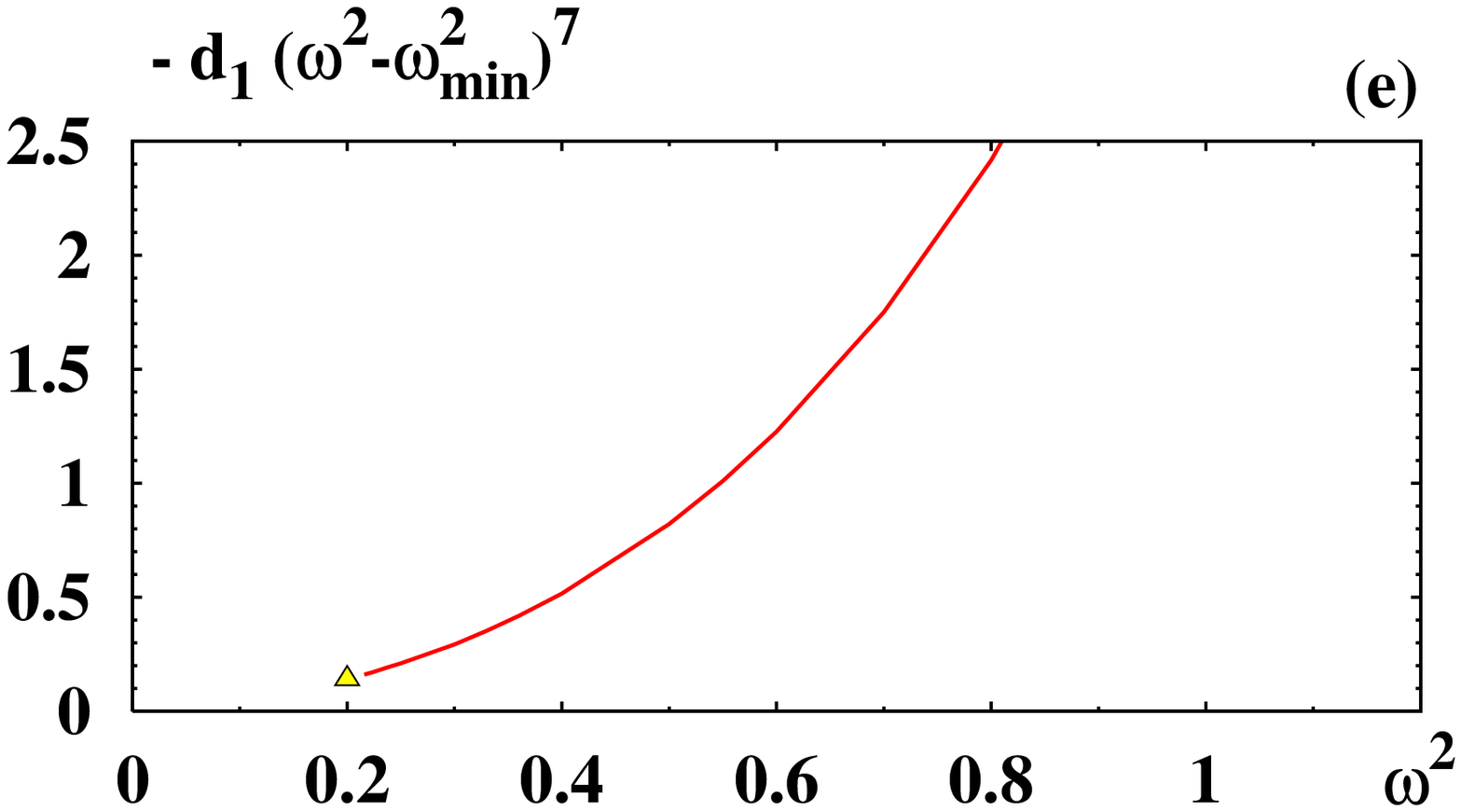}

\caption{\label{Fig-10:liq-drop-abs}
  $Q$-ball properties   as functions of $\omega^2$ 
  plotted in the form $X(\omega^2-\omega_{\rm min}^2)^N$.
  The respective quantities $X$, and their scaling powers $N$,
  written as the pairs $(X,N)$, are as follows:
  (a) $(\gamma,0)$,
  (b) $(\la r_s^2\ra,2)$,
  (c) $(Q,3)$,
  (d) $(M,3)$,
  (e) $(d_1,7)$.
  The triangles mark the analytical predictions from 
  Eqs.~(\ref{Eq:def-gamma-Coleman-II},
        \ref{Eq:rs2-liq-limit},
        \ref{Eq:Q-liq-limit},
        \ref{Eq:M-liq-limit},
        \ref{Eq:d1-liq-limit}).}

\vspace{-5mm}

\end{figure}

The surface energy $E_{\rm surf}$ diverges because it is proportional
to the surface area, which diverges for $\omega\to\omega_{\rm min}$.
Taking out carefully these divergences allows one to define 
the surface tension, cf.\ Eq.~(2.19) in \cite{Coleman:1985ki}, as
\ba   \!\!\!
      \lim\limits_{\omega\to\omega_{\rm min}} \!\gamma
     =\lim\limits_{\omega\to\omega_{\rm min}} 
      \int\limits_0^{\phi_0}\di\phi\;\sqrt{2\hat{U}}
      \, , \;\;\;
      \label{Eq:def-gamma-Coleman}\!\!\!
\ea
where $\hat{U}=V(\phi)-\frac12\omega^2\phi^2$.
Let us define
\be
   \varepsilon_{\rm min}=\sqrt{\omega^2-\omega_{\rm min}^2}>0\,.
\ee
With the substitution $\phi\to x = \phi^2$ we obtain
\be
   \int\limits_0^{\phi_0}\!\di\phi\;\sqrt{2\hat{U}}
   \equiv
   \frac{1}{2}\int\limits_0^{\phi_0^2(\varepsilon_{\rm min})}\!\!\!\di x\;
   \sqrt{2C\biggl(\frac{B}{2C}-x\biggr)^2 -\varepsilon_{\rm min}^2}
\ee
For $\varepsilon_{\rm min}\neq 0$ the integrand is complex.
Recalling that for $\varepsilon_{\rm min}\to 0$ we have
$\phi_0^2(\varepsilon_{\rm min})\to\phi_{\rm const}^2=B/(2C)$, 
see App.~\ref{App-A:trivial-solutions},
we obtain, for our parameters,
\be\label{Eq:def-gamma-Coleman-II}
      \lim\limits_{\omega\to\omega_{\rm min}} \!\gamma
      =  \frac{\sqrt{C}}{2\sqrt{2}}\,\biggl(\frac{B}{2C}\biggr)^2 
      = \frac{1}{2\sqrt{2}}\,,
\ee
which agrees with the numerical results, 
see Fig.~\ref{Fig-10:liq-drop-abs}a.

Next we want to determine the behavior of the mean square radius 
$\la r_s^2\ra$ in this limit. From Eq.~(\ref{Eq:constant-solutions})
we obtain for $p(0)$, Eq.~(\ref{Eq:pressure}), the behavior 
\be\label{Eq:p(0)-liq-limit}\;
        p(0) = \frac12\,\omega^2\phi_0^2-V(\phi_0) = \frac{B}{4C}\,
        \varepsilon^2_{\rm min} +{\cal O}(\varepsilon^4_{\rm min})\,,
\ee
which we have seen in Fig.~\ref{Fig-08:liq-drop}d
and derived alternatively in App.~\ref{App-B:bound-on-pressure}.
Combining this result with 
(\ref{Eq:prediction-liquid-drop-I},~\ref{Eq:def-gamma-Coleman-II}) 
yields for our potential
\be\label{Eq:rs2-liq-limit}
      \lim\limits_{\varepsilon_{\rm min}\to0} 
      \varepsilon_{\rm min}^4\,\la r_s^2\ra
      =\frac{B^2}{2C} = 2\,,
\ee
which is supported by the numerical results in
Fig.~\ref{Fig-10:liq-drop-abs}b.
Analogously we obtain
\ba\label{Eq:Q-liq-limit}\
      \lim\limits_{\varepsilon_{\rm min}\to0}\varepsilon_{\rm min}^6\,Q
      &=&\frac{\pi}{3\sqrt{2}}\;\frac{B^4}{C^{5/2}}\;\omega_{\rm \min}\;,\\
      \label{Eq:M-liq-limit}
      \lim\limits_{\varepsilon_{\rm min}\to0}\varepsilon_{\rm min}^6\,M
      &=&\frac{\pi}{3\sqrt{2}}\;\frac{B^4}{C^{5/2}}\;\omega_{\rm \min}^2\;,\\
      \label{Eq:d1-liq-limit}
      \lim\limits_{\varepsilon_{\rm min}\to0}\varepsilon_{\rm min}^{14}
      \,d_1
      &=&-\,\frac{\pi^2}{144}\;\frac{B^{10}}{C^6}\,\omega_{\rm \min}^2\;.
\ea
The numerical results in Figs.~\ref{Fig-10:liq-drop-abs}c--e fully 
support these conclusions. We see that among the quantities in 
(\ref{Eq:def-gamma-Coleman-II}--\ref{Eq:d1-liq-limit})
$d_1$ has the most rapid rise for $\omega\to\omega_{\rm min}$, which 
explains the observations in Fig.~\ref{Fig-05:ground-state-properties}. 
Combining (\ref{Eq:rs2-liq-limit},~\ref{Eq:M-liq-limit},~\ref{Eq:d1-liq-limit})
yields
\be   \label{Eq:d1-liq-limit-rescaled}
      \frac{d_1}{M^2\la r_s^2\ra} = -\,\frac14\;
      \frac{\varepsilon_{\rm min}^2}{\omega_{\rm \min}^2}\;+\;\dots\;,
\ee
where the dots indicate higher order terms. 
Eq.~(\ref{Eq:d1-liq-limit-rescaled}) explains the observation made 
in Fig.~\ref{Fig-06:d1-rescaled}, namely that $d_1$ measured in its 
``natural units'' vanishes in this limit. 

The liquid drop analogy was very successful. One could be tempted
to drive it further than we did it here, e.g., by giving the drop 
also a uniform charge distribution. The resulting repulsive forces 
ensure stability, and a virial theorem analog to 
(\ref{Eq:virial-theorem-III}) can be derived. 
But the microscopic details of the stabilizing dynamics are different 
from a $Q$-ball, and we will not pursue this analogy further.

\section{\boldmath The limit $\omega\to\omega_{\rm max}$ }
\label{Sec-5:limit-omega-to-max}

In this section we discuss the properties of $Q$-balls for
$\omega\to\omega_{\rm max}$. For certain potentials one obtains 
small and stable $Q$-balls in this so-called
``thick-wall'' limit \cite{Kusenko:1997ad}. 
But in our potential for $\omega^2>\omega_c^2\approx 1.9$
the $Q$-balls are unstable. For instance, the solution 
$\omega^2\approx 2.192$ with $Q=42$ and $M\approx 62.4$ 
can decay into 2 absolutely stable $Q$-balls corresponding
to $\omega^2\approx 1.223$ with $Q=21$ and $M\approx 28.5$,
or into 3 absolutely stable $Q$-balls corresponding
to $\omega^2\approx1.466$ with $Q=14$ and 
$M\approx 20.4$.\footnote{
   Here we content ourselves to state that the decays are possible 
   energetically, but we are not concerned about their dynamics.
   Notice also that in these examples integer charges were chosen. 
   But in general the charge $Q$ is not quantized.
   Also ``asymmetric'' decays into $Q$-balls of different charges 
   are possible, but then less energy is released.}

Finally, as $\omega\to\omega_{\rm max}$ in our potential, 
the solutions get more and more spread out, and approach from above 
$M\to m Q$ where $m=\omega_{\rm max}$ is the mass of the quanta, 
see Fig.~\ref{Fig-04:omega-abs}.
This means that the unstable $Q$-balls dissociate into a gas of 
free quanta, a ``$Q$-cloud'' \cite{Alford:1987vs}.

The aim of this section is to study analytically how $Q$-ball
properties behave for $\omega\to\omega_{\rm max}$.
The key for that is the large-$r$ asymptotics of $\phi(r)$
derived in Eq.~(\ref{Eq:asymp-large}) which shows that
as long as $\omega^2<\omega_{\rm max}^2$ the solutions $\phi(r)$ 
decay at large $r$ fast enough to ensure the convergence of the
integrals appearing in $M$, $Q$, or other properties.
Of course, the existence condition (\ref{Eq:condition-for-existence})
requires $\omega$ to be always smaller than $\omega_{\rm max}$. 
But we may study the scaling of $Q$-ball properties
as $\omega$ approaches $\omega_{\rm max}$ from below.

Let us define 
\be\label{Eq:phi-asymptotic}
    \phi_{\rm asymp}(r)=\frac{c_\infty}{r}\;
    e^{-\varepsilon_{\rm max} r}\,,\;\;\;
    \varepsilon_{\rm max}=\sqrt{\omega_{\rm max}^2-\omega^2}>0,
\ee
which is the leading term in the large-$r$ asymptotics of 
$\phi(r)$ in (\ref{Eq:asymp-large}). 
We  consider first the charge $Q$ in Eq.~(\ref{Eq:charge}) 
\ba
   Q &=& 4\pi\,\omega\int\limits_0^\infty{\di r}\,r^2\phi^2(r)\nonumber\\
     &\approx & \dots 
      +  4\pi\,\omega\int\limits_{\dots}^\infty{\di r}\,r^2\phi_{\rm asymp}^2(r)
    \nonumber\\
     &=& \dots 
      +  \frac{4\pi\,\omega\,c_{\infty}^2}{\varepsilon_{\rm max}}
         \int\limits_{\dots}^\infty \di x\; \exp(-2x)\;,
    \label{Eq:Q-substituted}
\ea
where in the second step we split the integral into an inner 
(indicated by the three dots) and an outer part. It is understood that
this decomposition is done at a sufficiently large radius $R$ such that 
$\phi(r)$ can be well approximated by its asymptotic form 
(\ref{Eq:phi-asymptotic}) for $r>R$. In the third step in
(\ref{Eq:Q-substituted}) we made the substitution 
$r\to x=\varepsilon_{\rm max}r$.

\begin{figure}[t!]

\vspace{-2mm}

\includegraphics[width=7.8cm]{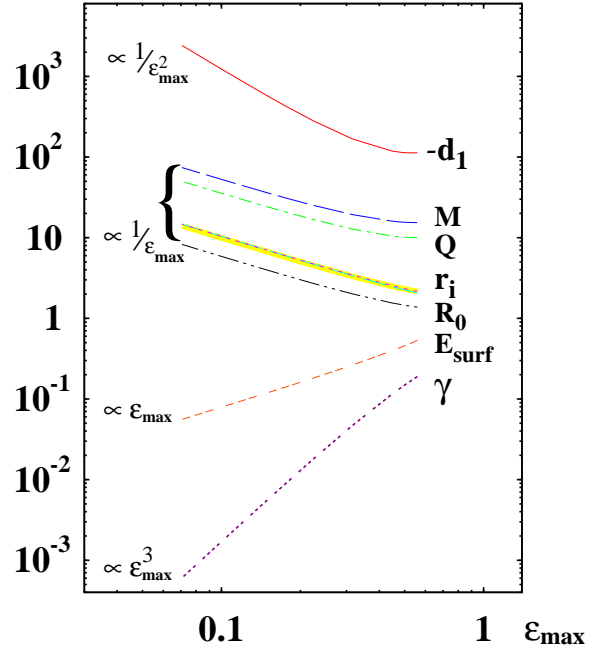}

\vspace{-2mm}
\caption{\label{Fig-11:Q-cloud-limit}
  The $Q$-ball properties $d_1$, $M$, $Q$, $R_0$, 
  $r_i=\la r^2_i\ra^{1/2}$ ($i=E,\,Q,\,s$), $E_{\rm surf}$, $\gamma$ 
  as functions of 
  $\varepsilon_{\rm max}=\sqrt{\omega^2_{\rm max}-\omega^2}$. 
  The region in the plot covers the range $1.8\le\omega^2\le 2.195$. 
  The~small-$\varepsilon_{\rm max}$ scaling of $d_1$, $M$, $Q$, 
  $\la r^2_Q\ra^{1/2}$ and $\la r^2_E\ra^{1/2}$ was predicted analytically 
  in  Eqs.~(\ref{Eq:Q-liq-limit-max}--\ref{Eq:r2-liq-limit-max}).}
\end{figure}

From Eq.~(\ref{Eq:Q-substituted}) we see what happens as 
$\varepsilon_{\rm max}$ decreases.
The inner part indicated by the three dots in 
(\ref{Eq:Q-substituted}) gives a finite contribution to $Q$,
but the outer contribution scales like $1/\varepsilon_{\rm max}$.
Thus, we expect that with decreasing $\varepsilon_{\rm max}$ 
the product $\varepsilon_{\rm max}Q \to {\rm const}$.
This method though does not allow us to determine the value of the 
constant. For that a more careful analysis is needed, which we will 
report elsewhere \cite{work-in-progress}.
But in this way we correctly predict that 
$Q\propto 1/\varepsilon_{\rm max}$ at small $\varepsilon_{\rm max}$, 
which is fully supported by the numerical results, see
Fig.~\ref{Fig-11:Q-cloud-limit}. 

Applying this method to other quantities we obtain the results 
summarized below
(all constants are positive and different in each case)
\ba   \label{Eq:Q-liq-limit-max}
      \lim\limits_{\varepsilon\to 0} \varepsilon_{\rm max}\,Q 
      &=& \;\;\;{\rm const},\\
      \label{Eq:M-liq-limit-max}
      \lim\limits_{\varepsilon\to 0} \varepsilon_{\rm max}\,M 
      &=& \;\;\; {\rm const},\\
      \label{Eq:d1-liq-limit-max}
      \lim\limits_{\varepsilon\to 0}\, \varepsilon^2_{\rm max}\,d_1\; 
      &=&-{\rm const},\\
      \label{Eq:r2-liq-limit-max}
      \lim\limits_{\varepsilon\to 0} \, \varepsilon^2_{\rm max}\,\la r^2_k\ra 
      &=&  \;\;\;{\rm const}, \;\; k = Q,\,E,\,s,\\
      \label{Eq:Esurf-liq-limit-max}
      \lim\limits_{\varepsilon\to 0} \varepsilon^{-1}_{\rm max}\,E_{\rm surf} 
      &=&  \;\;\; {\rm const}\, ,\label{Eq:cloud-limit-Esurf}\\
      \label{Eq:gamma-liq-limit-max}
      \lim\limits_{\varepsilon\to 0} \varepsilon^{-3}_{\rm max}\,\gamma 
      &=&  \;\;\; {\rm const}\, .
\ea
The predictions (\ref{Eq:Q-liq-limit-max}--\ref{Eq:gamma-liq-limit-max})
are fully supported by the numerical results as shown in 
Fig.~\ref{Fig-11:Q-cloud-limit}. 

Notice that the results 
(\ref{Eq:r2-liq-limit-max}--\ref{Eq:gamma-liq-limit-max})
for $E_{\rm surf}$, $\gamma$, $\la r_s^2\ra$ are numerical observations,
because our method cannot be applied to quantities vanishing with
$\varepsilon_{\rm max}\to0$.
In fact, for instance the scaling of the outer contribution in the integral 
(\ref{Eq:virial-theorem-Ib}) defining $E_{\rm surf}$ does imply
Eq.~(\ref{Eq:cloud-limit-Esurf}).
But our rough method would generically suggest that the contribution 
of the inner region scales like $\varepsilon_{\rm max}^0$ and dominates.
A more careful analysis is needed to prove the prediction 
(\ref{Eq:cloud-limit-Esurf}), see \cite{work-in-progress}.
The same reservations apply to the scaling behavior of $\gamma$ 
and $\la r_s^2\ra$, which are both connected to $E_{\rm surf}$
via Eq.~(\ref{Eq:def-gamma-rs}).

We observe numerically that the position $R_0$ where $p(r)$ changes
sign scales in the same way as the square roots of 
the mean square radii in (\ref{Eq:r2-liq-limit-max}). 
Thus, independently of whether we measure it in terms of $R_0$ or 
the $\la r_i^2\ra^{1/2}$, the size of the solutions
grows with $\varepsilon_{\rm max}\to 0$. 

The constant $d_1$ diverges as $1/\varepsilon_{\rm max}^2$ with 
decreasing $\varepsilon_{\rm max}$, see Fig.~\ref{Fig-11:Q-cloud-limit}. 
However, when measured in its natural units it actually
goes to zero as $d_1/(M^2\la r_s^2\ra) \propto \varepsilon_{\rm max}^2$
as we have seen previously in Fig.~\ref{Fig-06:d1-rescaled}. 

To summarize, as $\varepsilon_{\rm max}\to 0$ mass, charge and 
size of the $Q$-balls diverge as $1/\varepsilon_{\rm max}$, 
see (\ref{Eq:Q-liq-limit-max}, \ref{Eq:M-liq-limit-max},
\ref{Eq:r2-liq-limit-max}). Thus, the mean charge 
and energy densities, which are proportional to
$Q/({\rm size})^3$ and $M/({\rm size})^3$, 
vanish like $\varepsilon_{\rm max}^2$.  
Fig.~\ref{Fig-04:omega-abs} has shown that the $Q$-balls are unstable,
and their (positive) binding energy $M-m\,Q$ approaches zero (from above)
as $\varepsilon_{\rm max}\to 0$. Hence, in this limit we obtain a dilute 
gas of free $Q$-quanta as discussed in \cite{Alford:1987vs}.

\section{\boldmath  The sign of $d_1$}
\label{Sec-6:the-proof}

In this section we will show in several independent ways that $d_1$ 
is negative.
In Sec.~\ref{Sec-6.1:sign-of-d1-from-s} we will use for that the 
observation that for $Q$-balls $s(r)$ {\sl happens} to be positive 
for $0 < r < \infty$, and in 
Sec.~\ref{Sec-6.2:sign-of-d1-from-particle-interpretation} we will 
explain why $s(r)$ {\sl must} be positive.
In Sec.~\ref{Sec-6.3:sign-of-d1-from-p} we will prove that $d_1<0$ 
using arguments based on $p(r)$ and stability.

One may wonder why several proofs are needed.
Indeed, the EMT conservation dictates that
$s(r)$ and $p(r)$ are connected by the differential equation 
(\ref{Eq:diff-eq-s-p}), which is the origin of the equivalent 
presentations (\ref{Eq:def-d1-shear},~\ref{Eq:def-d1-pressure}) 
for $d_1$ in terms of $s(r)$ and $p(r)$ \cite{Goeke:2007fp},
and we have explicitly proven that our expressions for $s(r)$ and 
$p(r)$ satisfy (\ref{Eq:diff-eq-s-p}).
So, if one is able to conclude from $s(r)$ 
the sign of $d_1$, then it must be possible to draw the same conclusion 
also from $p(r)$. Therefore, at first glance it may seem sufficient to
conclude the sign of $d_1$ in one way, and below we will see
that for $Q$-balls it is much easier to use $s(r)$ for that.

However, concluding the sign of $d_1$ from $s(r)$ alone bears some 
danger, because from {\sl any} ``input'' $s(r)$ one obtains via
(\ref{Eq:diff-eq-s-p}) a pressure $p(r)$ which automatically\footnote{ 
  \label{Footnote:p-from-s-and-vice-versa}
  The differential equation (\ref{Eq:diff-eq-s-p}) allows one 
  to determine $p(r)$ from a given input function $s(r)$  
  only up to an integration constant. But the latter can be fixed 
  by demanding that for a well-localized finite-energy object 
  $p(r)\to0$ as $r\to\infty$. In a similar way one can determine
  $s(r)$ from a given input function $p(r)$.}
satisfies the stability condition (\ref{Eq:stability-condition})
\cite{Goeke:2007fp}. 
So one may well encounter an approach with $s(r)\ge 0$ and conclude 
$d_1<0$ {\sl without} being sure one really deals with a correct 
solution of the equations of motion and a true minimum of the 
energy functional. But
the other way round, the pressure is ultimately related to the 
issue of stability by Eq.~(\ref{Eq:stability-condition}), which we have 
shown to be equivalent to the virial theorem (\ref{Eq:virial-theorem-III}).
A proof that $d_1<0$ on the basis of $p(r)$ is therefore in general
on a much more solid ground.

\subsection{\boldmath Arguments based on $s(r)$ and inequalities}
\label{Sec-6.1:sign-of-d1-from-s}

In this section we will show that $d_1<0$ using
arguments based on the shear force distribution.
The argument is trivial and makes use of the observation
that manifestly $s(r)=\phi^\prime(r)^2\ge 0$ $\;\forall \,r$.

In Eq.~(\ref{Eq:def-d1-shear}) we have seen that $d_1$ is given by 
$(-\frac43\pi M)$ times the integral over $r^4s(r)$ over $r$ from 
zero to infinity. Since $s(r) \ge 0$ this immediately implies 
$d_1 \le 0$. This inequality can be improved by recalling that
$\phi^\prime(r)<0$ for $0 < r < \infty$, see Sec.~\ref{Sec-3a:Ueff}.
Therefore $d_1 < 0$ which completes the proof.

The fact that $s(r)\ge 0$ can be further explored to derive 
an inequality showing that $d_1$ must be negative. Using 
(\ref{Eq:thickness-of-wall},~\ref{Eq:def-mean-rs-radii}) we have
$d_1=-\frac{4\pi}3M\,\gamma\,\langle\langle r^4\rangle\rangle$
and $\la r_s^2\ra=\la\la r^2\ra\ra$ and can rewrite the constant
$d_1$ as
\be\label{Eq:d1-relation-with-wall-width}
     -\;\frac{d_1}{M^2\la r_s^2\ra} = \frac{M-\omega\,Q}{M}\,
     \Biggl(1+\biggl(\frac{\Delta r_s^2}{\la r_s^2\ra}\biggr)^{\!2}\Biggr)
     \;.\ee
Notice we implicitly benefited from the fact that $s(r)\ge 0$, 
when introducing the averages $\la\la r^n\ra\ra$
in (\ref{Eq:def-mean-rs-radii}). 
Next we explore that $E_{\rm surf}=\int\di^3r\,s(r)>0$, and
with $\omega\,Q>0$ we conclude from (\ref{Eq:M-omegaQ-Esurf}) 
that $0 < M-\omega\,Q<M$. Using the latter inequality in 
(\ref{Eq:d1-relation-with-wall-width}) finally implies that
\be\label{Eq:ineq-d1-fluctuations}
    0 <  
     -\;\frac{d_1}{M^2\la r_s^2\ra} < 
    1+\biggl(\frac{\Delta r_s^2}{\la r_s^2\ra}\biggr)^{\!2} \;.
\ee
This proves that $(-\,d_1)>0$. As a byproduct
Eq.~(\ref{Eq:ineq-d1-fluctuations}) provides 
also an upper bound on  $(-d_1)$ 
but in terms of $\Delta r_s^2/\la r_s^2\ra$. At this point it is not 
obvious whether this quantity is bound from above, though numerically we 
observe this to be the case in Fig~\ref{Fig-09:liq-drop-width-of-wall}.
That $\Delta r_s^2/\la r_s^2\ra$ is indeed bound from above 
will be shown in \cite{work-in-progress}.

For completeness let us mention the following more useful upper 
bound on the magnitude of $(-d_1)$. 
The staring point is Eq.~(\ref{Eq:d1-from-properties}) where we 
neglect the positive quantity 
$\omega\,Q\,M\,\la r^2_Q\ra=M\,\omega^2\int\di^3r\,r^2\phi(r)^2$.
This yields the bound
\be\label{Eq:d1-from-properties-IIB}
    - \,\frac{d_1}{M^2\la r_E^2\ra} < \frac59 \;,
\ee
which is satisfied by the numerical results, see
Fig.~\ref{Fig-06:d1-rescaled}.
We checked that this is the strongest inequality one can derive 
involving $\la r_E^2\ra$ as length scale.
The inequality (\ref{Eq:d1-from-properties-IIB}) is atrractive
because it provides an upper bound on $(-d_1)$ in ``its natural units'' 
solely in terms of quantities related to the energy density $T_{00}(r)$.

\subsection{Arguments based on the particle interpretation}
\label{Sec-6.2:sign-of-d1-from-particle-interpretation}

In the previous section we explored the observation that 
$s(r)$ {\sl happens} to be positive for $0<r<\infty$
for $Q$-balls. 
Here we will show this {\sl must} be the case.

For that we use the particle interpretation picture 
\cite{Coleman:1985ki} discussed in Sec.~\ref{Sec-3a:Ueff}.
The Newtonian equation (\ref{Eq:Newtonian-eom}) describing
the motion of a unit mass particle moving in the effective potential 
$U_{\rm eff}=\frac12\omega^2\,x^2-V(x)$ under the friction 
$F_{\rm fric}=-\frac2t\,\mbox{\it\. x}(t)$ follows from the 
Langrange-function $L(\mbox{\it\. x},x)$ and Rayleigh's 
dissipation function ${\cal F}(\mbox{\it\. x})$,
\be
     L(\mbox{\it\. x},x) = \frac12\,\mbox{\it\. x}(t)^2-U_{\rm eff}(x),\;\;
     {\cal F}(\mbox{\it\. x}) = \frac1t\,\mbox{\it\. x}(t)^2 \;,
\ee
according to
\be
     \frac{{\rm d}\,}{{\rm d} t}\biggl(\frac{\partial L}{\partial\mbox{\it\. x}}\biggr)
     -\frac{\partial L}{\partial x} = 
     - \;\frac{\partial\cal F}{\partial\mbox{\it\. x}} \;.
\ee
The physical meaning of Rayleigh's dissipation function 
${\cal F}(\mbox{\it\. x})$ is that it describes the rate at which
the system dissipates its energy $E$ due to the frictional force,
namely
\be
    \frac{{\rm d} E}{{\rm d} t} = \frac{{\rm d}\,}{{\rm d} t}\biggl(
    \partial\mbox{\it\. x}\,\frac{\partial L}{\partial\mbox{\it\. x}}-L\biggr)
    = -2\,{\cal F} \le 0\;\;\;\forall\;t\,,
\ee
which must be negative because the system {\sl dissipates} energy.
This means that ${\cal F}(\mbox{\it\. x})\ge0$ $\;\forall\,t$.

If we recall that $x(t)$ and $t$ in the particle interpretation picture
correspond to $\phi(r)$ and $r$, we instantly see that 
${\cal F}(\mbox{\it\. x})$ corresponds to $\frac{1}{r}\,s(r)$.
Since ${\cal F}(\mbox{\it\. x})\ge 0$ this proves that
the distribution of shear forces $s(r)\ge 0$.

\subsection{\boldmath Arguments based on the pressure $p(r)$}
\label{Sec-6.3:sign-of-d1-from-p}

In this section we will prove that $d_1$ is negative, 
basing our arguments on the pressure distribution. 

Let us first demonstrate that for $\omega$ satisfying 
the existence condition (\ref{Eq:condition-for-existence}) the 
pressure is positive at small $r$, and negative at large $r$.
In Sec.~\ref{Sec-3a:Ueff} we have proven $p(0)>0$, and for reasons 
of continuity $p(r)>0$ also in some vicinity of the origin.
At large-$r$ we derive from (\ref{Eq:asymp-large}) the 
following asymptotics for the pressure 
\be
    p(r) = -\,(\omega_{\rm max}^2-\omega^2)\,
           \frac{2\,c_\infty^2}{3\,r^2}\,\exp\left(
           -2\,r\,\sqrt{\omega_{\rm max}^2-\omega^2}\right)
            + \dots
\ee
where the dots indicate subleading terms.
Clearly, for all $\omega$ satisfying the existence condition
(\ref{Eq:condition-for-existence}), the pressure is negative 
at large $r$. To summarize, we have 
\ba
 &&    p(r) > 0 \;\;\;\mbox{for small $r$,}\nonumber\\
 &&    p(r) < 0 \;\;\;\mbox{for large $r$.} \label{Eq:pressure-small-large-r}
\ea
This implies that $p(r)$ must change the sign an odd number 
of times. Of course, $p(r)$ must change sign at least
to comply with the stability 
condition (\ref{Eq:stability-condition}). From physical point
of view, we expect $p(r)$ to be positive in the~center 
(which implies repulsive forces directed towards outside)
and negative outside (attractive forces towards inside),
as we derived in (\ref{Eq:pressure-small-large-r}).
A stable solution arises when the repulsive and attractive
forces exactly balance each other according to 
(\ref{Eq:stability-condition}).
This physically intuitive pattern was observed 
also in soliton models of the nucleon
\cite{Goeke:2007fp,Goeke:2007fq,Cebulla:2007ei,Kim:new}.

\begin{figure}[t]
\includegraphics[width=8cm]{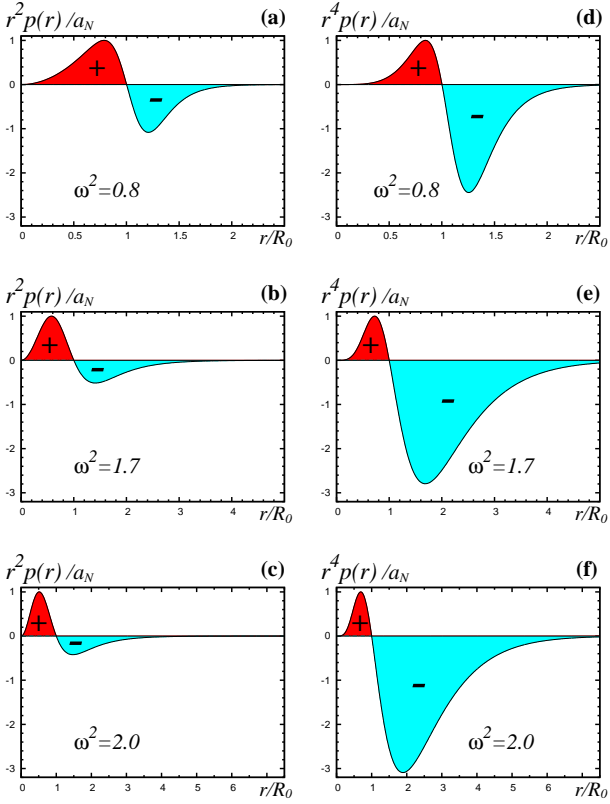}

\vspace{-2mm}
\caption{\label{Fig-12:p-r2-vs-r4}
$r^2p(r)$ and $r^4p(r)$ as functions of $r$ for selected $\omega$. 
For better comparison, $r$ is given in units of the radius $R_0$ 
where $p(r)$ changes sign, and the normalization factors $a_N$ 
are such that the curves reach unity at their global maxima.
The left (right) panel shows the integrand of the stability
condition (the integrand of $d_1$, Eq.~(\ref{Eq:def-d1-pressure}),
up the prefactor $5\pi M$). The figure illustrates why $d_1$ is negative.
Integrating the curves in the left panel yields zero due to the
stability condition. Weighting the curves
by an additional factor of $r^2$ and integrating then yields a negative
result for $d_1$, see right panel.}
\end{figure}

For a ground state one may expect the pressure
distribution to change sign only once, see 
Fig.~\ref{Fig-02:ground-state-densities}d.
If we {\sl assume} $p(r)$ to change sign one and only one time,
this immediately implies that $d_1$ is negative.
Fig.~\ref{Fig-12:p-r2-vs-r4} illustrates the argument.
The left panel of Fig.~\ref{Fig-12:p-r2-vs-r4} visualizes 
the stability condition (\ref{Eq:stability-condition}): 
the shaded areas above and below the x-axis are equal 
and exactly compensate each other. 
Thus, due to the stability condition (\ref{Eq:stability-condition})
we have
\ba
     \int\limits_0^{R_0}     \di r\;\underbrace{\;R_0^2\,r^2\,}_{\displaystyle > r^4} \,p(r) 
 &=&-\int\limits_{R_0}^\infty\di r\;\underbrace{\;R_0^2\,r^2\,}_{\displaystyle < r^4} \,p(r)\nonumber\\
     \Rightarrow \;\;\;\;\;
     \int\limits_0^{R_0}     \di r\;r^4 p(r) 
 &<&-\int\limits_{R_0}^\infty\di r\;r^4 p(r)
\label{Eq:proving-sign-of-d1-from-pressure-ground-state}
\ea
which means $\int_0^\infty\di r\;r^4 p(r)<0$,
and $d_1$ must be negative, as can be seen in the
right panel of Fig.~\ref{Fig-12:p-r2-vs-r4}.

If we knew $p(r)$ has one zero only,
the proof that $d_1<0$ would be complete here.
It is intuitive to assume that the pressure
distribution of a ground state changes sign 
only once according to
(\ref{Eq:pressure-small-large-r}). However,
here we will provide a general argument
based on stability which is valid not only
for ground states. 

For that we need the following lemma. For any solution
of the $Q$-ball equations of motion we have
\be\label{Eq:proving-d1-sign-from-p-1.0}
     \int\limits_0^R\di r\;r^2\,p(r) > 0 
     \;\;\;\;\;\mbox{for}\;\;\;\;\; 
     0 < R < \infty. 
\ee
To prove (\ref{Eq:proving-d1-sign-from-p-1.0}) we make use 
of the result (\ref{Eq:proving-stability-1.3})
derived in Sec.~\ref{Subsec-3b:conservation-of-EMT-and-Qballs}
and explore the particle interpretation picture of the 
$Q$-ball equations of motion. For that we notice that
$p(r)+\frac23\,s(r)$, the right-hand-side of 
(\ref{Eq:proving-stability-1.3}), is positive $\forall \;r<\infty$
because
\be\label{Eq:proving-d1-sign-from-p-1.4}
     \bigg(p(r)+\frac23\;s(r)\biggr)
     = \underbrace{\,\frac12\;\phi^\prime(r)^2\,}_{\displaystyle  E_{\rm kin}}
     \;+\;
     \underbrace{\frac12\omega^2\phi(r)^2-V(\phi)}_{\displaystyle U_{\rm eff}}
     \:.
\ee
In other words, $p(r)+\frac23\,s(r)$ corresponds to 
the total, kinetic plus potential, energy of the 
particle at a given time $t$ (with $t\leftrightarrow r$).
The total energy of the particle must be larger than zero 
$\forall\;t<\infty$, because at any finite time $t$ the particle still 
needs to do work against the friction in order to arrive, after 
infinite time, at the origin $x=0$ (with $x\leftrightarrow \phi$). 
This means that the integral in (\ref{Eq:proving-stability-1.3})
is positive and proves Eq.~(\ref{Eq:proving-d1-sign-from-p-1.0}).

To understand this result intuitively, we remark that if the 
integral (\ref{Eq:proving-d1-sign-from-p-1.0}) could become zero at 
some finite $R$, then the fields would stabilize themselves 
in a subinterval $r\in [0,R]$. Then, setting the fields outside 
that interval to zero, would yield a stable solution with lower mass 
(because $T_{00}(r)\ge 0$ also in the omitted region), 
in contradiction to the expectation that a given set of initial 
value data leads to a unique minimum of the action.

We have now all ingredients for the proof that $d_1<0$
based on pressure and stability arguments, namely
(\ref{Eq:pressure-small-large-r}) and
(\ref{Eq:proving-d1-sign-from-p-1.0}).
The proof is as follows.

Eq.~(\ref{Eq:pressure-small-large-r}) means $p(r)$ must change 
sign an odd number $N$ of times. Let $R_i$ with 
$i=1,\,\dots\,,\,N$ denote the radii where this happens with 
$0<R_1<R_2<\dots <R_N<\infty$. Notice that we do not include 
points where $p(r)$ could have zeros without changing sign.

The stability condition (\ref{Eq:stability-condition}) can then
be written as
\be\label{Eq:proving-d1-sign-from-p-2.0}
    \int\limits_{0}^{R_1}\!\di r\;r^2p(r) +
    \int\limits_{R_1}^{R_2}\!\di r\;r^2p(r) + \dots +
    \int\limits_{R_N}^\infty\!\di r\;r^2p(r) = 0\;.
\ee
By construction $p(r)\ge 0$ in the first, third, $\dots$ integrals,
and $p(r)\le 0$ in the second, fourth, $\dots$ last integrals.
We will now replace each of the terms in 
(\ref{Eq:proving-d1-sign-from-p-2.0}) by a smaller term,
and show in this way that $\int_0^\infty\!\di r\;r^4p(r) < 0$.

{\bf Step 1.}
We consider the first 2 terms in (\ref{Eq:proving-d1-sign-from-p-2.0}). 
In the first (second) term $p(r)$ is positive (negative). Therefore
\ba \label{Eq:proving-d1-sign-from-p-2.1}
    \int\limits_0^{R_1}\!\di r\;r^2p(r) 
    &\ge& \frac1{R_1^2}\int\limits_0^{R_1}\!\di r\;r^4p(r)\,,\nonumber\\
    \int\limits_{R_1}^{R_2}\!\di r\;r^2p(r) 
    &\ge& \frac1{R_1^2}\int\limits_{R_1}^{R_2}\!\di r\;r^4p(r)\,.
\ea
Adding up the 2 inequalities in (\ref{Eq:proving-d1-sign-from-p-2.1}) 
we obtain
\ba \label{Eq:proving-d1-sign-from-p-2.2}
    \int\limits_0^{R_2}\!\di r\;r^2p(r) 
    \ge \frac1{R_1^2}\int\limits_0^{R_2}\!\di r\;r^4p(r)\;.
\ea
If there is only one change of sign, then we take 
the limit $R_2\to\infty$ and recover the situation of 
Eq.~(\ref{Eq:proving-sign-of-d1-from-pressure-ground-state}),
and our proof is complete here.
If $p(r)$ changes sign more than once, i.e.\ $N\ge 3$,
then we have to continue our proof and include further contributions
in step 2.

{\bf Step 2.}
Notice that $\int_0^{R_2}\!\di r\;r^4p(r)>0$ for $R_2<\infty$ due to
(\ref{Eq:proving-d1-sign-from-p-1.0}). Moreover $R_1<R_3<\infty$.
Therefore
\ba \label{Eq:proving-d1-sign-from-p-2.3}
    \int\limits_0^{R_2}\!\di r\;r^2p(r) 
    \ge \frac1{R_1^2}\int\limits_0^{R_2}\!\di r\;r^4p(r)
    > \frac1{R_3^2}\int\limits_0^{R_2}\!\di r\;r^4p(r)\,.
\ea
In the next two intervals $p(r)\ge 0$ for $r\in[R_2,R_3]$, 
and $p(r)\le 0$ for $r\in[R_3,R_4]$. Therefore,
in analogy to (\ref{Eq:proving-d1-sign-from-p-2.1}), 
\ba \label{Eq:proving-d1-sign-from-p-2.4}
    \int\limits_{R_2}^{R_3}\!\di r\;r^2p(r) 
    &\ge& \frac1{R_3^2}
    \int\limits_{R_2}^{R_3}\!\di r\;r^4p(r)\,,\nonumber\\
    \int\limits_{R_3}^{R_4}\!\di r\;r^2p(r) 
    &\ge& \frac1{R_3^2}
    \int\limits_{R_3}^{R_4}\!\di r\;r^4p(r)\,.
\ea
Combining (\ref{Eq:proving-d1-sign-from-p-2.3}) and the results in
(\ref{Eq:proving-d1-sign-from-p-2.4}) we obtain
\ba \label{Eq:proving-d1-sign-from-p-2.5}
    \int\limits_{0}^{R_4}\!\di r\;r^2p(r) 
    & > & \frac1{R_3^2}
    \int\limits_{0}^{R_4}\!\di r\;r^4p(r)\;.
\ea
If $p(r)$ changes sign exactly $N=3$ times, then we take 
in (\ref{Eq:proving-d1-sign-from-p-2.5}) the limit
$R_4\to\infty$ and our proof is completed here.
If $p(r)$ changes sign more often, then we repeat 
step 2.

{\bf\boldmath Last step.} If $p(r)$ changes the sign 
$N=2k+1$ times (recall that $N$ is odd, and $R_N<\infty$), 
then we repeat successively the $2^{\rm nd}$ step $k$-times, 
until we arrive at 
\ba \label{Eq:proving-d1-sign-from-p-2.6}
    \int\limits_{0}^\infty\!\di r\;r^2p(r) 
    & > & \frac1{R_N^2}
    \int\limits_{0}^\infty\!\di r\;r^4p(r)\;.
\ea
Now, the first integral vanishes due to the stability condition 
(\ref{Eq:stability-condition}), and using the definition 
(\ref{Eq:def-d1-pressure}) giving $d_1$ in terms of the pressure, 
we conclude from (\ref{Eq:proving-d1-sign-from-p-2.6}) the desired 
result that $d_1$ must be negative.

\subsection{\boldmath Stability and $d_1$}
\label{Sec-6D:stability-vs-d1}

In the above section we have proven that the stability condition
(\ref{Eq:stability-condition}) and the properties of the pressure
at small and asymptotically large $r$ unambiguously yield to a
negative sign of the constant $d_1$. Therefore, clearly if we
have a stable object then $d_1<0$.

The inverse, however, is not true in general. A negative $d_1$ 
does not imply the object is stable. Fig.~\ref{Fig-12:p-r2-vs-r4}
illustrates this point. 
The solution in Fig.~\ref{Fig-12:p-r2-vs-r4}a for 
$\omega^2=0.8<\omega_{\rm abs}^2$ is absolutely stable. 
The $Q$-ball solution in Fig.~\ref{Fig-12:p-r2-vs-r4}b for 
$\omega_{\rm abs}^2<\omega^2=1.7<\omega_c^2$ is metastable.
The solution in Fig.~\ref{Fig-12:p-r2-vs-r4}c for 
$\omega_c^2<\omega^2=2.0$ is unstable.
But in all 3 cases $d_1$ is negative, as the 
Figs.~\ref{Fig-12:p-r2-vs-r4}d--e illustrate.

The point is that all solutions for 
$\omega_{\rm min}<\omega<\omega_{\rm max}$ correspond to minima 
of the energy functional. Therefore all solutions satisfy the stability 
condition (\ref{Eq:stability-condition}), or equivalently the 
virial theorem (\ref{Eq:virial-theorem-III}), as well as all
conditions required to prove that they have a negative $d_1$.
But only for $\omega_{\rm min}<\omega<\omega_{\rm abs}$ we deal 
with global minima of the action and absolutely stable $Q$-balls.

\section{Conclusions}
\label{Sec-7:conclusions}

We have presented a study of $Q$-balls in a scalar field theory with 
U(1) symmetry, and investigated the properties of $Q$-balls as functions 
of the angular velocity $\omega$. While $Q$-ball stability was
studied in literature before
\cite{Volkov:2002aj,Gleiser:2005iq,Gani:2007bx,Sakai:2007ft,Tsumagari:2008bv},
to best of our knowledge this is the 
first study in which this issue is addressed from the point of view 
of the EMT, and the constant $d_1$. All solutions 
presented in this work were exact solutions of the equations of motion. 
Particular focus was put on the behavior of $Q$-ball properties
for $\omega$ approaching the boundaries $\omega_{\rm min,max}$ 
of the region in which solutions exist.

For $\omega\to\omega_{\rm min}$ the $Q$-balls occupy increasingly large 
volumes filled with $Q$-ball matter of nearly constant density
\cite{Coleman:1985ki}.
We have shown that in this limit the $Q$-ball properties follow the 
predictions of the liquid drop picture. Certain $Q$-ball properties such 
as charge $Q$, mass $M$, mean square radii, and $d_1$ 
diverge as $\omega\to\omega_{\rm min}$. We derived analytically the 
limits for these and other properties, which are
fully supported by our numerical results.

In the opposite limit $\omega\to\omega_{\rm max}$ the solutions 
become unstable and approach the ``$Q$-cloud limit'' \cite{Alford:1987vs}. 
Also in this limit some properties diverge. We derived analytically 
the scaling behavior of these quantities as $\omega$ approaches 
$\omega_{\rm max}$.
Further results will be reported in \cite{work-in-progress}.
It is remarkable that, among all properties we studied, $d_1$ 
diverges most strongly as $\omega\to\omega_{\rm min,max}$.

The conservation of the EMT implies among others the stability 
(von-Laue-) condition \cite{Polyakov:2002yz,von-Laue} 
which states that the pressure must satisfy
$\int_0^\infty \di r\:r^2p(r)=0$, which we have proven explicitly
in two independent ways. One of the proofs is equivalent to the virial
theorem.

The central result of this work is that the constant $d_1$ is 
strictly negative for all $Q$-ball solutions, 
\be\label{Eq:sign-d1}
     d_1 < 0 \;,
\ee
for which we have provided 2 explicit analytical proofs.
One proof involved the relation of $d_1$ to $s(r)$, and made use of 
the Newtonian particle interpretation of the $Q$-ball equations 
of motion \cite{Coleman:1985ki} in which the shear force distribution 
$s(r)$ is related to the Rayleigh dissipation function describing the 
frictional forces.
Since the Newtonian system dissipates energy due to
friction, $s(r)$ must be positive for $0<r<\infty$. 
This implies $d_1<0$. 

The other proof explored the relation of $d_1$ to $p(r)$, and made 
use of stability arguments. We have shown, using the equations of 
motion, that $p(r)$ is positive in the center of the $Q$-ball 
(which corresponds to repulsion) and negative at large $r$
(which corresponds to attraction). 
This means $p(r)$ must change the sign an odd number of times,
and we have formulated a general proof valid for any $Q$-ball
solution with a pressure with an arbitrary number of zeros.
We observed that for ground states $p(r)$ changes sign
only once, but for radial excitations of $Q$-balls one encounters
more complex structures \cite{work-in-progress}.

The proof of (\ref{Eq:sign-d1}) based on $p(r)$ elucidates that for 
$Q$-balls the negative sign of $d_1$ is a consequence of stability.
This was conjectured, but could not be proven rigorously, also in 
other soliton approaches 
\cite{Goeke:2007fp,Goeke:2007fq,Cebulla:2007ei}.

The last important insight of our study is that stability implies 
$d_1<0$, but the opposite is not true. A negative $d_1$ does 
not necessarily mean the object is stable. In fact, our
proofs hold equally for stable, metastable and unstable
$Q$-balls. 
To guarantee a negative $d_1$ both proofs require that
we deal with the exact solution of the equations of motion,
i.e.\ with a minimum of the action, be it global or local.
But absolute stability requires that the minimum of the action is global.
Thus, stability is a {\sl sufficient} but {\sl not necessary}
criterion for $d_1$ to be negative. 

An interesting open question is how quantum fluctuations 
\cite{Graham:2001hr} modify the picture of $d_1$ and the
proof $d_1<0$. It would be also interesting to see
how $d_1$ is altered for $Q$-balls coupled to fermionic 
fields, which allows them to 'evaporate' \cite{Cohen:1986ct}.
The ultimate goal would be to generalize the proofs given in 
this work to quantum field theories, and to apply them to the 
description of hadrons. So far, in all theoretical studies $d_1$ 
was always found negative, for pions \cite{Polyakov:1999gs}, nucleons 
\cite{Petrov:1998kf,Goeke:2007fp,Goeke:2007fq,Cebulla:2007ei,Kim:new}.
and nuclei \cite{Polyakov:2002yz,Liuti:2005gi,Guzey:2005ba}.  
Also lattice QCD calculations yield a negative quark contribution to 
the $d_1$ of nucleon, though the gluon contribution and hence the 
total $d_1$ are not yet known \cite{Mathur:1999uf}.
First experimental results are compatible with $d_1$ being negative
\cite{Ellinghaus:2002bq} but this observation is not yet conclusive
\cite{Belitsky:2001ns}, and future data will provide further insights 
\cite{CLAS-proposal}.

\

\noindent{\bf Acknowledgments.}
We thank Gerald Dunne and Alex Kovner for helpful discussions.
M.~M.~is grateful to the Universities of Connecticut and
Heidelberg for support during initial stages of this work. 
The work was partly supported by DOE contract DE-AC05-06OR23177.

\appendix

\section{\boldmath Stationary solutions}
\label{App-A:trivial-solutions}

In this Appendix we discuss stationary solutions of the equations 
of motion (\ref{Eq:eom},~\ref{Eq:boundary-conditions}) of the type 
$\phi(r)=\phi_{\rm const}$ $\forall r$. Though they do not obey the 
boundary condition (\ref{Eq:boundary-conditions}) at infinity, 
these solutions are nevertheless of interest.

If $\phi(r)=\phi_{\rm const}$ the boundary conditions at $r=0$
in (\ref{Eq:boundary-conditions}) hold, and the equation of motion
(\ref{Eq:eom}) becomes
\ba\label{Eq-App:triv-sol-I}
&&  \hspace{-5mm} 
    \omega^2\phi_{\rm const} - V^\prime(\phi_{\rm const}) \\
&&  = 
    \biggl(\omega^2-2A+4B\phi_{\rm const}^2-6C\phi_{\rm const}^4\biggr)
    \phi_{\rm const} = 0. \nonumber
\ea
The trivial solution $\phi_{\rm const} = 0$ describes the vacuum.
Further (for $\omega_{\rm min} \le \omega \le \omega_{\rm max}$
always real) solutions are
\be\label{Eq-App:triv-sol-II}
    \phi_{\rm const}^2 = \frac{B}{3\,C}\,\pm\,
    \biggl(\frac{B^2}{9\,C^2}+\frac{\omega^2-2A}{6C}\biggr)^{1/2} .
\ee
Because ${\cal L}$ in (\ref{Eq:Lagrangian}) is symmetric under 
$\phi\to -\phi$ it is sufficient to focus on the non-negative 
solutions. 

In the particle interpretation picture the two positive stationary 
solutions in (\ref{Eq-App:triv-sol-II}) have the following meaning.
The solution with the minus-sign in (\ref{Eq-App:triv-sol-II})
corresponds to the situation that the particle is at $t=0$
precisely in the local minimum of the effective potential
in Fig.~\ref{Fig-01:eff-pot} and will stay there forever.
The solution with the plus-sign in (\ref{Eq-App:triv-sol-II})
corresponds to the (not stable) situation that the particle is at $t=0$
precisely in the global maximum of the effective potential
in Fig.~\ref{Fig-01:eff-pot} and will stay there forever.

As $\omega\to\omega_{\rm min}$ we obtain $\phi_{\rm const}^2= B/(6C)$
from (\ref{Eq-App:triv-sol-II}), which corresponds to the (not 
interesting for us) situation where the particle stays forever 
in the minimum of $U_{\rm eff}$, and the (much more interesting) 
solution
\be
\label{Eq-App:triv-sol-IVa}
   \phi_{\rm const}^2 \to \frac{B}{2\,C} = 1\;\;\;\mbox{as}\;\;\;
   \omega\to\omega_{\rm min}\,, 
\ee
which corresponds to the situation with the particle placed at the 
maximum of $U_{\rm eff}(\phi)$ which for $\omega$ approaching 
$\omega_{\rm min}$ from above is just above zero. 
This situation is of interest, because as
$\omega\to\omega_{\rm min}$ the particle has to be placed very 
close to this maximum of $U_{\rm eff}(\phi)$, and ``wait'' there long 
enough such that its small initial potential energy $U_{\rm eff}(\phi)$ 
is sufficient to overcome the time-dependent friction which decreases 
with time \cite{Coleman:1985ki}.

As $\omega\to\omega_{\rm max}$ we obtain from
(\ref{Eq-App:triv-sol-II}) the solutions $\phi_{\rm const}^2=0$ 
and $\phi_{\rm const}^2=(2B)/(3C)$. When $\omega=\omega_{\rm max}$
the effective potential does not dip below zero at all, i.e.\ 
it is not possible to release the particle from any $\phi_0>0$
so it would stop in the origin \cite{Coleman:1985ki}. 
The only solution is  $\phi_{\rm const}^2=0$. However, from
(\ref{Eq-App:triv-sol-II}) that solution develops in the limit 
$\omega\to\omega_{\rm max}$ from the minimum of $U_{\rm eff}$ 
which is below zero.
But for any regular solution with $\omega<\omega_{\rm max}$
the potential at the starting point  $U_{\rm eff}(\phi_0)>0$.
Therefore, the stationary solution $\phi_{\rm const}^2=0$
obtained here is not continuously connected to the limiting
value for $\phi_0$ stated in Eq.~(\ref{Eq:phi-in-limiting-cases}).

\section{Bounds on the pressure}
\label{App-B:bound-on-pressure}

The pressure at the origin is just $p(0)=U_{\rm eff}(\phi_0)$
and this is positive, Eq.~(\ref{Eq:pressure-positive-at-small-r}),
because for any regular solution the effective potential of the
particle at the starting point must be positive. In this Appendix 
we will show that the pressure is also bound from above. From 
(\ref{Eq:boundary-conditions},~\ref{Eq:pressure}) we obtain
\be\label{Eq:p(0)-II}
     p(0) = \frac12\,\omega^2\phi_0^2
     \biggl(1-\frac{2\,V(\phi_0)}{\omega^2\,\phi_0^2}\biggr)\,.
\ee
Using (\ref{Eq:condition-for-existence-2}) we notice that 
$\forall\,\phi(r)$ (including $\phi(r)$ at $r=0$)
\be\label{Eq:p(0)-III}
    \frac{2\,V(\phi)}{\phi^2} \ge 
    \min\limits_\phi \biggl[\frac{2\,V(\phi)}{\phi^2}\biggr]
    \equiv \omega_{\rm min}^2\,.
\ee
Inserting (\ref{Eq:p(0)-III}) in (\ref{Eq:p(0)-II}) and including
also the lower bound from
Eq.~(\ref{Eq:pressure-positive-at-small-r}) we obtain
\be\label{Eq:App-upper-bound-p(0)}
    0 < p(0) \le \frac12\,(\omega^2-\omega_{\rm min}^2)\,\phi_0^2 \;.
\ee
An important application of the upper bound in 
(\ref{Eq:App-upper-bound-p(0)}) is that it allows to 
verify independently the result for $p(0)$ in the limit 
$\omega\to\omega_{\rm min}$ quoted in 
(\ref{Eq:prediction-liquid-drop-densities}). 
We remark that the upper bound in (\ref{Eq:App-upper-bound-p(0)})
is for $\omega_{\rm min}<\omega<\omega_{\rm max}$ always a real 
inequality and saturated only in the limit $\omega\to\omega_{\rm min}$.

\section{\boldmath Generating functional for $\la\la r^n\ra\ra$}
\label{App-C:alternative-expressions-gamma-rs2}

In Sec.~\ref{Sec-4:limit-omega-to-min} we defined 
the $\la\la r^n\ra\ra$ in (\ref{Eq:def-mean-rs-radii}), which 
allowed us to express compactly $\la r_s^2\ra=\la\la r^2\ra\ra$
and the measure for the wall width 
$(\Delta r_s^2)^2=\la\la r^4\ra\ra-\la\la r^2\ra\ra^2$.
Another interesting application, if we continue to negative $n$, is 
$p(0)=2\,\gamma\:\la\la r^{-1}\ra\ra$. This allows us to express
the result obtained in (\ref{Eq:prediction-liquid-drop-I}) as
\be\label{Eq:C1}
     \lim\limits_{\omega\to\omega_{\rm min}}\, 
     \la\la r^{-1}\ra\ra\;\la\la r^2\ra\ra^{1/2} = 1.
\ee
The result in Eq.~(\ref{Eq:C1}) can be understood and interpreted
by recalling that $s(r)\to\gamma\,\delta(r-R)$ in the liquid drop 
limit which is equivalent to $\omega\to\omega_{\rm min}$, see
Sec.~\ref{Sec-4:limit-omega-to-min}.

The positivity of the shear forces, 
which was crucial in Secs.~\ref{Sec-6.1:sign-of-d1-from-s} 
and \ref{Sec-6.2:sign-of-d1-from-particle-interpretation}, 
allows us to introduce the functional
\be
      F(\lambda) = \int\limits_0^\infty\di r\;s(r)\;{\rm exp}(-\lambda r^2)\,
\ee
which is a generating functional for $\la\la r^n\ra\ra$
for even $n$
\be
      F(\lambda) = F(0)\sum\limits_{n=0}^\infty
      \frac{(-1)^n}{n!} \;\la\la r^{2n}\ra\ra\;\lambda^n
\ee

The surface tension is just $\gamma=F(0)$, 
the surface energy is $E_{\rm surf}=-4\pi F^\prime(0)$ and 
$d_1=-\frac{4\pi}{3}\,M\,F^{\prime\prime}(0)$. The mean square radius 
of the shear forces, and the measure of the wall width $\Delta r_s^2$ 
can be expressed as
\ba
     \la r_s^2\ra &=& 
     -\;\biggl[\frac{\partial}{\partial\lambda}\,{\rm log}\;F(\lambda)
        \biggr]_{\lambda=0}
     \,,\\
     (\Delta r_s^2)^2 &=& \phantom{-\;}
     \biggl[\frac{\partial^2}{\partial\lambda^2}\,{\rm log}\;F(\lambda)
     \biggr]_{\lambda=0}\,.
\ea

We remark that each integral over $s(r)$ can be traded for an integral 
over $p(r)$ by exploring the differential equation (\ref{Eq:diff-eq-s-p}).
For instance, $\gamma = \frac34\,\int_0^\infty\di r\;p(r)$ and
\ba
    \la r_s^2\ra = -\;6\;
    \frac{\int_0^\infty\di r\;r^2p(r)\,{\rm log}\,r}{
          \int_0^\infty\di r\;p(r)\;}\, .
\ea
These and further relations were derived in \cite{Goeke:2007fp}
and can be used as cross checks for the numerics.


\end{document}